%% file: PALM_ArXiv.tex
\documentclass[12pt]{article} 
\usepackage{color}
\usepackage{graphicx}
\usepackage{amssymb}
\usepackage{amsmath}
\usepackage{natbib}
\usepackage{setspace}
\usepackage{standalone}
\usepackage{fullpage}
\usepackage{amsthm}
\usepackage{amsfonts}
\usepackage{amsmath}
\usepackage{natbib}
\usepackage{hyperref}
\usepackage{algorithmic}
\usepackage[boxed]{algorithm}
\usepackage{graphicx}
\usepackage{booktabs}

\begin{document}

	\title{Precision aggregated local models}
	\author{Adam M.~Edwards\thanks{Corresponding author: Department of Statistics, Virginia Tech,
			Hutcheson Hall, 250 Drillfield Drive
			Blacksburg, VA 24061, USA;
			\href{mailto:adaedwar@vt.edu}{\tt adaedwar@vt.edu}}
		\and Robert B.~Gramacy\thanks{Department of Statistics, Virginia Tech}
	}
	\date{}

	\maketitle

	\begin{abstract} Large scale Gaussian process (GP) regression is infeasible
		for larger data sets due to cubic scaling of flops and quadratic
		storage involved in working with covariance matrices. Remedies in
		recent literature focus on divide-and-conquer, e.g., partitioning into
		sub-problems and inducing functional (and thus computational)
		independence.  Such approximations can be speedy, accurate, and
		sometimes even more flexible than an ordinary GPs.  However, a big
		downside is loss of continuity at partition boundaries. Modern methods
		like local approximate GPs (LAGPs) imply effectively infinite
		partitioning and are thus pathologically good and bad in this regard.
		Model averaging, an alternative to divide-and-conquer, can maintain
		absolute continuity but often over-smooths, diminishing accuracy.
		Here we propose putting LAGP-like methods into a local experts-like
		framework, blending partition-based speed with model-averaging
		continuity, as a flagship example of what we call precision aggregated
		local models (PALM).  Using $K$ LAGPs, each selecting $n$ from $N$
		total data pairs, we illustrate a scheme that is at most cubic in $n$,
		quadratic in $K$, and linear in $N$, drastically reducing
		computational and storage demands. Extensive empirical illustration
		shows how PALM is at least as accurate as LAGP, can be much faster in
		terms of speed, and furnishes continuous predictive surfaces.
		Finally, we propose sequential updating scheme which greedily refines
		a PALM predictor up to a computational budget.

	\end{abstract}

	\section{Introduction} \label{sec:intro}

	Gaussian Processs (GPs) are popular for nonparametric modeling of non-linear
	functions, serving as surrogates for computer experiments
	\citep[e.g.,][]{sack:welc:mitc:wynn:1989,santner2018design,gramacy2020surrogates},
	models of spatially referenced data \citep{cressie:1993}, and as machine
	learners \citep{rasmu:will:2006}.  Conceptually a GP is very simple,
	assuming that all response values are multivariate normal (MVN) with a
	correlation function dictated by inverse (often Euclidean) distance in the
	input space.  While there are many ways to define the mean and covariance
	structure for GPs, here we use a simple default common in surrogate modeling
	and machine learning.  Given $N$ training data pairs $(X_N, Y_N)$, where
	$X_N$ is comprised $d$-variate row vectors $x_i^\top$ and $Y_N$ is an
	$N$-vector of scalars, take $Y_N \sim
	\mathcal{N}_N(0, \tau^2 (C_N + \eta \mathbb{I}_N))$, where the $(i,j)$
	coordinates of $K_N = C_N + \eta \mathbb{I}_N$ follow the so-called
	separable Gaussian kernel 
	\[ 
	K_{\theta, \eta}(x_i, x_j) =
	\exp\left\{-\sum_{\ell=1}^{d}\frac{\left(x_{i\ell}-x_{j\ell}\right)^2}{\theta_\ell}\right\}
	+ \eta \mathbb{I}_{\{i=j\}},
	\] 
	and $\theta = (\theta_1, \dots,
	\theta_d)$ is the collection of {\em lengthscale} hyperparameters determining the
	strength of correlation by distance in each direction.  The MVN covariance
	is augmented by a so-called {\em nugget} hyperparameter $\eta$, to allow
	smoothing over noisy data.  Above, $\mathbb{I}_{\{i=j\}}$ is the Kroneker
	delta function, returning unity only when input indices match identically.
	We shall begin by taking a deterministic computer surrogate modeling
	perspective and simplify with $\eta =\epsilon$. In this context $\epsilon$
	is sometimes referred to as the {\em jitter} \citep{neal1998regression}.

	GPs are popular because the MVN structure can be extended to new predictive
	locations $x$, where closed form conditioning yields that $Y(x) \mid (Y_N,
	X_N)$ is also Gaussian.  Given settings of hyperparameters, moments of that
	distribution are given as
	\begin{align}
		\hat{\mu}(x) &= k_N^\top(x) K_N^{-1} Y_N \label{eq:mux} & \mbox{and} && 
		\hat{\sigma}^2(x) &= \tau^2 (1 + \eta - k_N^\top(x) K_N^{-1} k_N (x)), 
	\end{align}
	where $k_N(x)$ is an $N$-dimensional column vector containing
	$k_{\theta}(x_i, x)$, for $i = 1, \dots, n$.  Note that $\eta$ is dropped
	from the subscript in $k_\theta$ because the Kroneker delta would evaluate
	to zero. Predictive surfaces under these equations have many desirable
	properties such as smoothness, and appropriate out-of-sample coverage. It
	can be shown that $\hat{\mu}(x)$ is a best linear unbiased predictor (BLUP),
	minimizing mean-squared prediction error among the class of models specified
	by the hyperparametrized GP kernel \citep[see, e.g.,][Chapter
	3.2]{santner2018design}.  Empirically, GP predictions are hard to beat
	out-of-sample.  Inference for $\theta$ and $\eta$  proceeds via ordinary MVN
	log likelihoods through $K_N^{-1}$ and $|K_N|$, perhaps leveraging closed
	form derivatives. Newton schemes work well \citep[see, e.g.,][Chapter
	5.2]{gramacy2020surrogates}.

	The big downside for GPs is evident in those equations above.  Prediction and
	likelihood-based inference relies on decomposing a potentially large matrix $N
	\times N$ matrix $K_N$, for $K_N^{-1}$ and $|K_N|$. Storage requirements are
	clearly quadratic in $N$, and the common Cholesky-based decomposition,
	yielding both inverse and determinant simultaneously, is cubic
	-- although slight improvements are available with fancier linear algebra.
	Without tuned libraries such as Intel MKL, most modern workstations can
	cope with GPs trained on at best $N \approx 5000$.  With MKL, the limit 
	is about $N \approx 10000$ \citep[][Appendix A]{gramacy2020surrogates}.

	Work in all three GP communities -- surrogate modeling, geostatistics, and
	machine learning -- has been feverish of late, in part as an attempt to
	cope with the ever increasing training data sizes $N$ produced by modern
	application and instrumentation.  A representative, but certainly not
	exhaustive, list across communities includes the following:
	\citet{snelson:ghahr:2006, haaland:qian:2012,gramacy:polson:2011,
	cressie:joh:2008, kaufman:etal:2012, sang:huang:2012,
	nychka:wykle:royle:2002, qc:rasmu:2005, furrer:genton:nychka:2006,
	katzfuss2017multi, katzfuss2018general}.  Here our focus is on partition
	models, which have enjoyed considerable success in the surrogate modeling
	literature, for example as based on treed partitioning
	\citep{gramacy2008bayesian} and Voronoi tesselation
	\citep{kim2005analyzing,rushdi_vps:_2016}.  Although these approaches
	considerably expanded upon the state-of-the-art in terms of both fidelity
	and computation, some have taken to more aggressive divide-and-conquer to
	handle even larger datasets.  Examples involving partitioning include
	Bayesian additive regression trees \citep[BART,
	][]{chipman2010bart,chipman2012sequential} and local approximate Gaussian
	processes \citep[LAGP,][]{gramacy2015local,gramacy2014massively}. Such
	predictors are highly accurate, fast, and their divide-and-conquer nature
	enable vast parallelization.  A downside to partition-based schemes,
	however, is that they result in discontinuous predictive surfaces
	-- pathologically so in the cases of these latter more aggressive schemes.
	Lack of smoothness is not only aesthetically inferior, it also suggests
	that predictability is being left on the table when the underlying dynamics
	are inherently smooth.

	In this paper we propose massaging divide-and-conquer GP regression into an
	aggregation framework in order to smooth over pathological discontinuity,
	guaranteeing a continuous surface, while at the same time providing a
	computationally {\em more} efficient predictor.  Our focus is on ``smoothing
	LAGP'' although the idea, in principle, can be extended to any predictor
	furnishing Gaussian predictive equations. Essentially, we propose using the
	local model's own predictive (inverse) variance (aka, precision) as a means
	of patching together and smoothing over functionally independent local
	predictors.  We call the framework PALM, for {\em precision aggregated local
	models}.  We begin in Section \ref{sec:partagg} by developing a framework
	that bridges partition/divide-and-conquer regression with model averaged
	alternatives, taking the opportunity to connect to recent methods in the
	literature which leverage such schemes effectively, and those which target
	smoothing over hard partition boundaries.

	Section \ref{sec:palm} introduces the PALM framework in generality, but
	ultimately emphasizes its application with LAGP base models, or ``local experts'',
	and the details required for effective implementation. 
	We illustrate how a limited number of LAGP experts, positioned in a
	space-filling manner, may be aggregated via precision to yield a fast global
	predictor which is smooth, and at least as accurate as the ordinary {\em
	transductive} \citep{vapnik2013nature} approach recommended by its
	creators \citet{gramacy2015local}: apply LAGP exhaustively and independently
	on each element of a testing set.  Section \ref{sec:seq} considers how local
	experts can be allocated sequentially, in a boosting-like framework to
	regions of the input space/testing set that are harder to predict.  Greedy
	selection allows for a more efficient use of limited computational resources
	compared to the default space-filling approach. Before concluding with
	remarks in Section \ref{sec:discuss}, Section \ref{sec:empirical} augments
	with empirical work extending the proof-of-concept illustrations provided in
	earlier sections.

	\section{Unified partition and aggregation}
	\label{sec:partagg}

	Partition models, such as based on trees or Voronoi tesselation,
	divide the data $\mathcal{D}$ geographically into mutually exclusive regions
	$\mathcal{D}_1, \dots, \mathcal{D}_K$ where $\mathcal{D}_k \cap
	\mathcal{D}_j = \emptyset$ for all $k \neq j$, and $\cup_{k=1}^K
	\mathcal{D}_k = \mathcal{D}$.  Inference for the splitting structure
	depends upon a goodness-of-fit criteria paired with choice of model for each
	$\mathcal{D}_i$, and a good search scheme.  This divide-and-conquer nature
	has two benefits.  One is that fitting smaller models is often cheaper,
	computationally, than fitting one big model. Nowhere is this more true than
	with GPs, requiring cubic scaling in flops to decompose large matrices.
	Another is modeling fidelity.  By fitting different models in different
	parts of the input space, the overall fit is more dynamic than its base
	version. For GPs, which usually exhibit stationary dynamics, independent
	modeling and fitting of hyperparameters ($\theta$ and $\eta$) provides a
	nonstationary flavor. Nonetheless, the underlying predictor remains within
	the GP class though an implicit sparse--block covariance structure.

	Another way to see this -- one which suits our proposed methodology and
	developments coming shortly -- is to represent the partitioned
	predictor as a weighted average.  Suppose we wish to predict at testing
	site $x$.  Let $w_k(x) = \delta(x \in \mathcal{D}_k)$ indicate
	whether or not $x$ is in partition element $\mathcal{D}_k$.  Then the
	partitioned predictor, using GP predictive notation (\ref{eq:mux})
	applied separately for each index $k$ via training data $(X_k, Y_k) \in
	\mathcal{D}_k$, may be represented as 
	\begin{align} 
		\hat{\mu}(x) &=
		\sum_{k = 1}^{K} \hat{\mu}_k(x) w_k(x)  & \mbox{and} && 
		\hat{\sigma}^2(x) &= \sum_{k=1}^{K} \sum_{j=1}^{K}
		\hat{\sigma}_{kj}(x) w_k(x) w_j(x). \label{eq:partvar}
	\end{align} 
	Above, $\hat{\sigma}_{kj}\left(x\right)$ represents the
	covariance between the predictors in partitions $k$ and $j$, and
	$\hat{\sigma}_{kk}(x) \equiv \hat{\sigma}_k^2(x)$.  Since the
	$\mathcal{D}_k$ are mutually exclusive, only one component of the sums 
	in Eq.~(\ref{eq:partvar}) receive any
	weight.  Each set of predictive equations is itself Gaussian, arising
	from a GP for $k=1,\dots,K$, so the weighted sum is also Gaussian and
	thus a GP.

\begin{figure}[ht!] 
	\centering 
	\includegraphics[width = .5\textwidth]{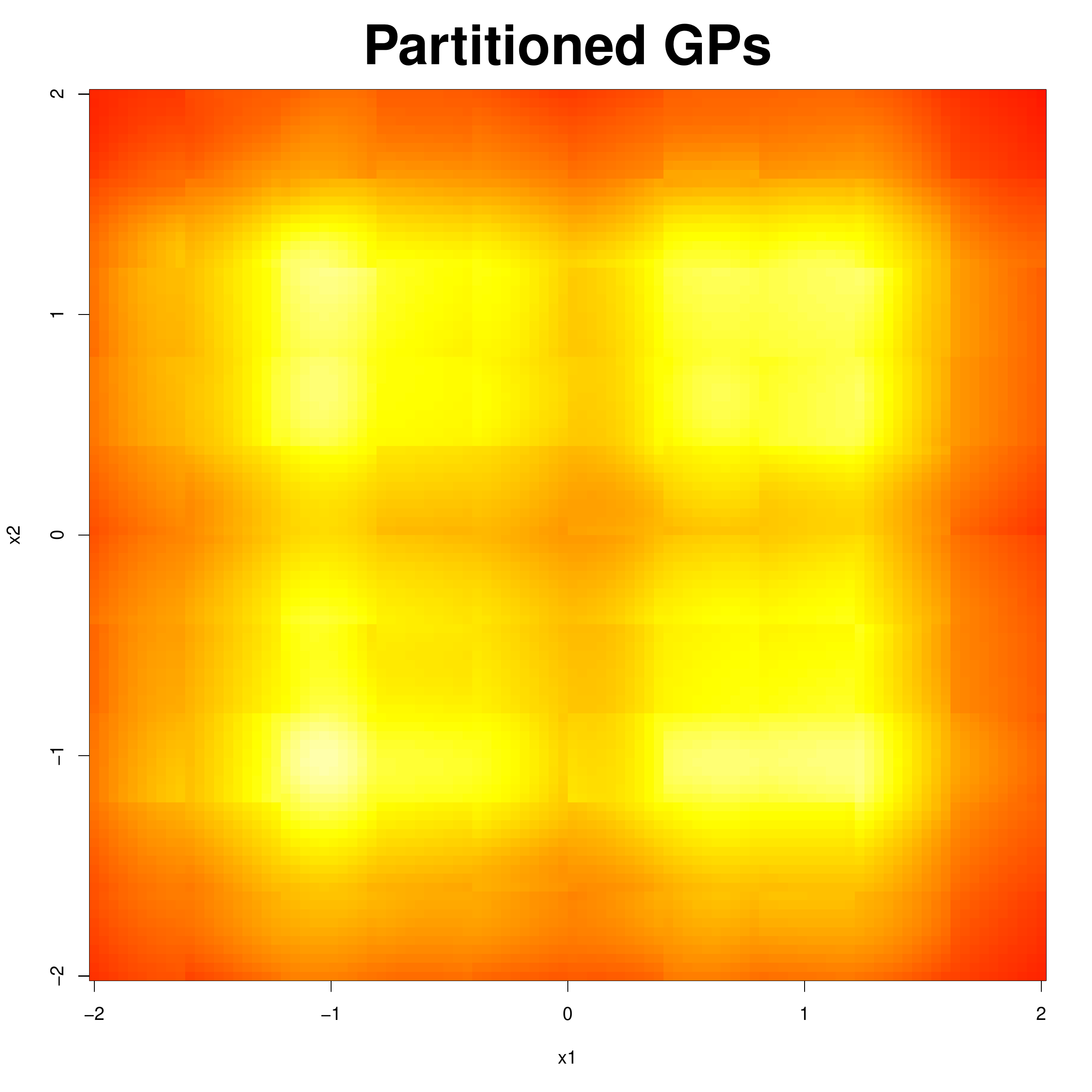} 
	\caption{Predictive mean obtained from uniform-partition GP fit to Herbie's
	tooth data.}
	\label{fig:part}
	\end{figure}

	To illustrate, consider  {\em Herbie's tooth} \citep{lee2011optimization}.
	For scalar inputs $z$, define
	\begin{equation} 
	g(z) = \exp\left(-(z-1)^2\right) +
	\exp\left(-0.8(z+1)^2\right) - 0.05\sin\left(8(z+0.1)\right).
	\label{eq:herbie} 
	\end{equation} 
	For inputs $x$ with $m$ coordinates $x_1,\dots,x_d$, the response is $f(x) =
	-\prod_{j=1}^d g(x_j)$.  Here the default $d=2$ case is chosen for ease of
	visualization.  Consider a training set composed of a $100 \times 100$
	uniform grid in $[-2,2]^2$ and commensurately sized ($101 \times 101$), but
	slightly shifted out-of-sample testing set.  In Figure \ref{fig:part}, a
	regular $K=100$-element partition is created, and GPs are fit independently
	to the training data residing in each element of the partition.  Full GPs
	are all but intractable on data of this size, requiring hours to fit on an
	ordinary workstation.\footnote{...  using ordinary linear algebra packages}
	Through divide-and-conquer, fitting each 100-run element in serial takes
	around two minutes total, and potentially much faster in parallel.

	From the figure, a downside to the  partition model is readily evident: lack
	of continuity at boundaries, which at times can be extreme.  Attempts to
	remedy these are not without their limitations. Some seem limited to, or
	have only been successfully illustrated in, two-dimensional input spaces,
	and can only guarantee continuity in the mean surface, while variances
	remains discontinuous \citep{park_domain_2011,park_efficient_2016}. {\em
	Patchwork kriging} \citep{park_patchwork_2017} can guarantee point-wise
	continuity at reference locations in both surfaces, but is discontinuous
	along the full manifold of partition boundaries. As the number of
	continuity--inducing reference points increases, it will converge to a fully
	continuous model.  Perhaps the biggest downside of approaches attempting to
	``stitch'' boundaries together is that it is not immediately obvious whether
	resulting predictor is also a GP.

	\subsection{Extreme partitioning and re-smoothing}

	A local approximate Gaussian process \citep[LAGP,][]{gramacy2015local} is a
	purely predictive process that builds a new $n \ll N$-sized model for each
	desired predictive location. In this way it is similar to the $n$-nearest
	neighbor (NN), although \citeauthor{gramacy2015local} show that when using
	GP prediction, NNs can be a suboptimal conditioning set.  Review of
	methodological and computational details of LAGP would be a distraction
	here.  See, e.g., \citet{gramacy_lagp:_2016} for details the {\tt laGP}
	package for {\sf R} on CRAN. Intuitively, the method exploits the rapidly
	decreasing influence of training points on predictive mean and variance as
	they increase in distance from the predictive location of interest. GP
	prediction based on a handful of, say $n=50$ local training sites is fast,
	highly parallelizable (over a vast predictive set), accurate, and
	yields a limited nonstationary effect not unlike partition methods.  A form
	of ``infinite partitioning'' is happening implicitly in the input space
	$\mathcal{D}$ space from the point of view of the testing set, which makes
	LAGP an example of {\em transductive learning} \citep{vapnik2013nature}, as
	opposed to the usual {\em inductive} sort.  It fits into the form of
	Eq.~(\ref{eq:partvar}) when enumerating all $K$ subsets
	$X_N$ which are of size $n$ and giving only positive weight to those which are
	``near'' to $x$.  Although that is a daunting enterprise to think about, its
	implementation is rather straightforward, and compact in terms of
	storage. Since LAGP finds a new model for each prediction, the storage
	cost of the model only grows with the size of the original data, instead of
	$\mathcal{O}\left(N_k^2\right)$ like most partition models.

	Despite many advantages, LAGP does have drawbacks -- many of
	which may be evident from the narrative above.  First, it does not
	create any permanent model.  Each time a new prediction is desired, a
	new fit must be calculated.  This leaves room for similar methods with a
	permanent model to drastically undercut, trading off space (to store
	the permanent model) for time. It does not avoid the continuity
	concerns experienced by other partitioning based schemes.  Instead, they
	are grossly exacerbated: discontinuities are everywhere, although their
	precise nature may sometimes be difficult to detect with the naked eye.  

	\begin{figure}[ht!]
		\centering
		\includegraphics[width = .5\textwidth]{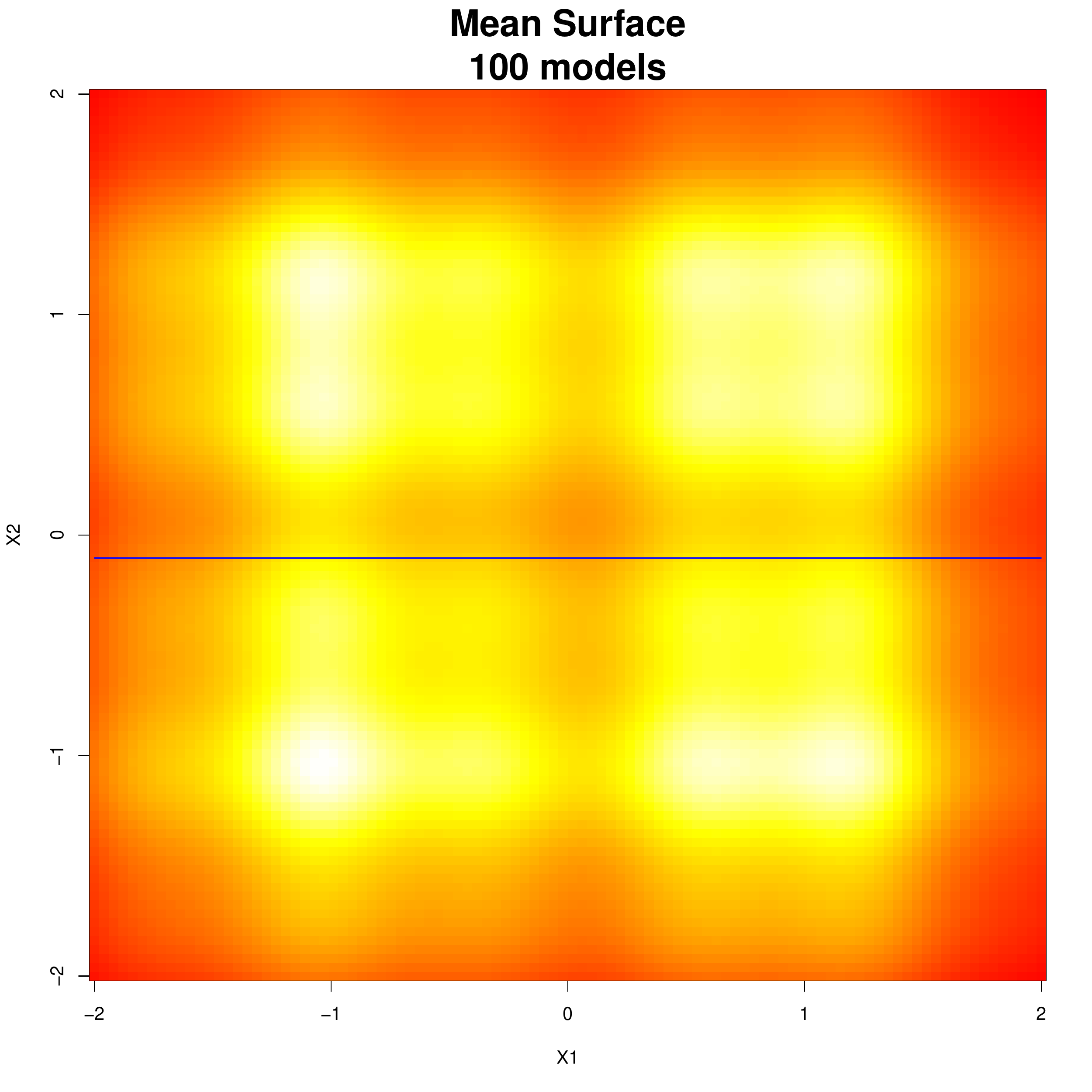}%
		\includegraphics[width = .5\textwidth]{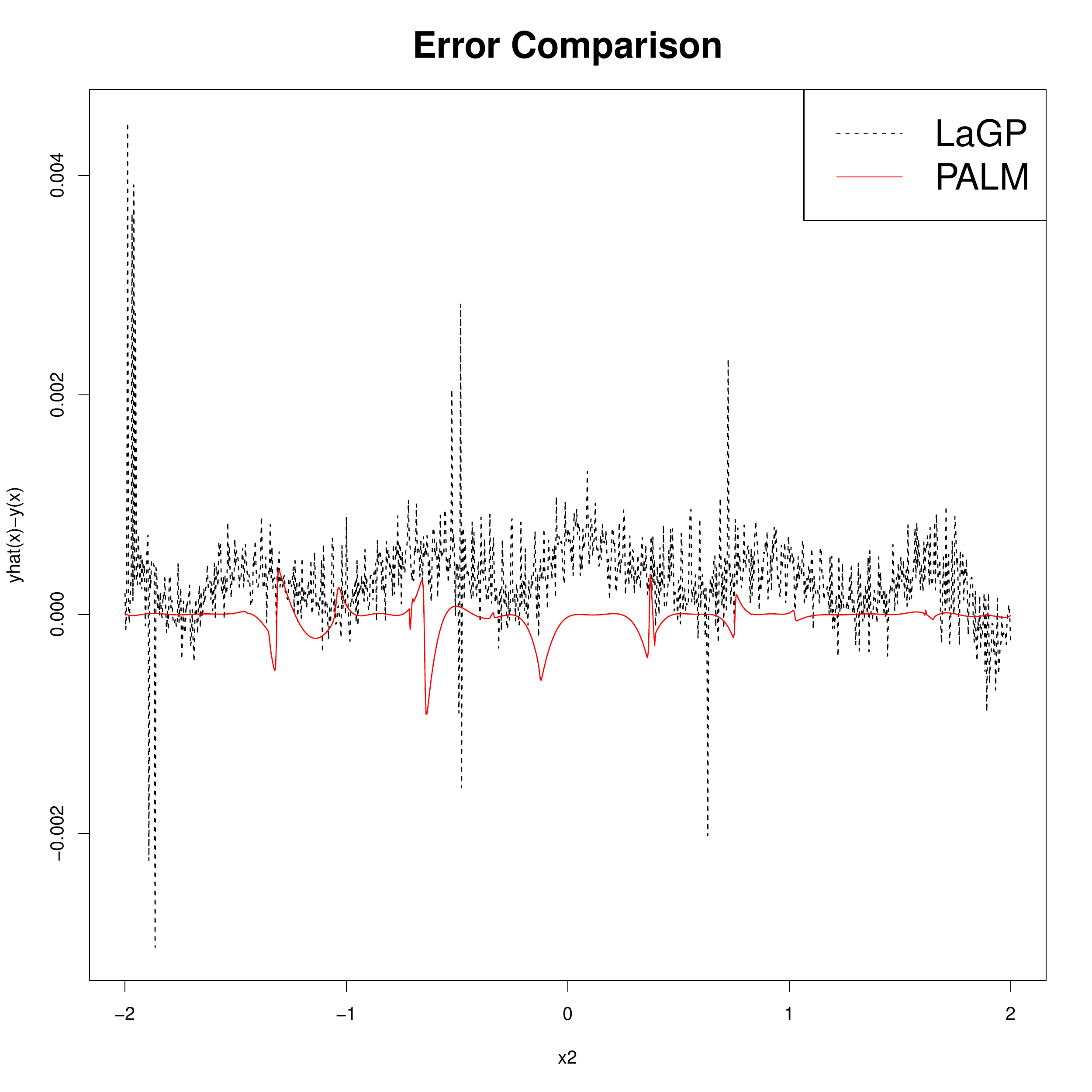}
		\vspace{-0.75cm} 
		\caption{Herbie's tooth fits: LAGP on
		left; bias (right) in a slice through LAGP measured
		against the truth and a PALM competitor.}
	\label{fig:error} \end{figure}

	Figure \ref{fig:error} shows an LAGP fit on Herbie's tooth data. The left
	panel mirrors Figure \ref{fig:part}, showing a predictive mean surface from
	LAGP using the defaults in the {\tt laGP} package which uses a neighborhood
	size of $n=50$ and greedy local selection by a method called {\em active
	learning Cohn} \citep[ALC;][]{cohn1994neural}.  The right panel in the
	figure provides a higher-resolution view via a slice defined by fixing $x_2$
	at $-0.104$, as indicated by the horizontal line on the left panel.  Rather
	than showing predictive means on the right, bias $\hat{y}(x) - y(x)$ is
	plotted instead as a means of ``zooming in''.  Focus for now on the dashed
	gray line, corresponding to LAGP.  Notice that LAGP's relationship to the
	true response, which is itself smooth, is pathalogically nonsmooth. Although
	the overall magnitude of the bias is quite small, at worst 0.006 in absolute
	value, there are isolated exceedingly poor predictions.  Although many
	biases are much smaller than that, being less than 0.001 in absolute value,
	there is a systematic (almost sinusoidal) patten in these out-of-sample
	residuals. The red curve, which corresponds to our proposed PALM method,
	introduced shortly, suggests that both issues -- smoothness and
	predictibility left on the table
	-- could be resolved with a modest amount of aggregating, inducing
	dependence between otherwise functionally independent local fits.

	\subsection{Model averaging} \label{sec:avg}

	So far we have only considered boolean weights in Eq.~(\ref{eq:partvar}).
	What about more general $w(\cdot)$ satisfying
	$\sum_{k=1}^{K}w_k\left(x\right) = 1$?  The result is a (true) model
	averaging predictor, representing a potentially smoother alternative to
	partition models.  Such setups have been combined with GPs in order to
	combat the computational burden of fitting a single large GP
	\citep{tresp_bayesian_2000}. Rather than partitioning, components being
	averaged can involve overlapping domains or random subsetting of the data,
	both with and without replacement.   A common weighting scheme for these
	models is $w_k(x) \propto \phi_k(x)$, where $\phi_k(x) \equiv
	1/\hat{\sigma}_k^2(x)$ is the predicted precision at $x$ from model $k$.
	Such a choice is optimal if the predictors indexed by $k$ are independent
	and unbiased \citep{cochran_combination_1954}. The unbiased assumption is
	often violated, e.g., GPs which are biased toward the prior
	mean, but precision weighting is nontheless a default.

	A bigger issue is independence. Because model averaging schemes abandon
	the use of indicator weights, and moreover may use overlapping data
	subsets, they must deal with multiple model covariance calculations
	$\sigma_{jk}(x)$ or risk inducing further bias into the meta-predictor
	(\ref{eq:partvar}).  Model averaging measures rely
	on the fact that predictions from component models are independent as
	long as they have learned the same function.  Functional independence
	of global models can be expressed probabilistically as $P(X_i \mid
	X_{-i}, f) = P(X_i \mid f)$ where $f$ is the global function.  While true functional independence cannot be guaranteed, most model averaging schemes are able to show $P(X_i \mid X_{-i}, f) \approx P(X_i \mid f)$. With
	this approximation we can use $\sigma_{jk}(x) = 0$ for all $i \neq j$
	since, if they are learning the same function, the separate models are
	independent.  Similar to the partition model, this reduces the variance
	calculation to $\hat{\sigma}^2\left(x\right) =
	\sum_{k=1}^{K}\hat{\sigma}_k^2\left(x\right)w_k^2\left(x\right)$.  The quality of this variance estimate is dependent on the approximation to functional independence.

\begin{figure}[ht!] 
		\centering
		\includegraphics[width = .5\textwidth]{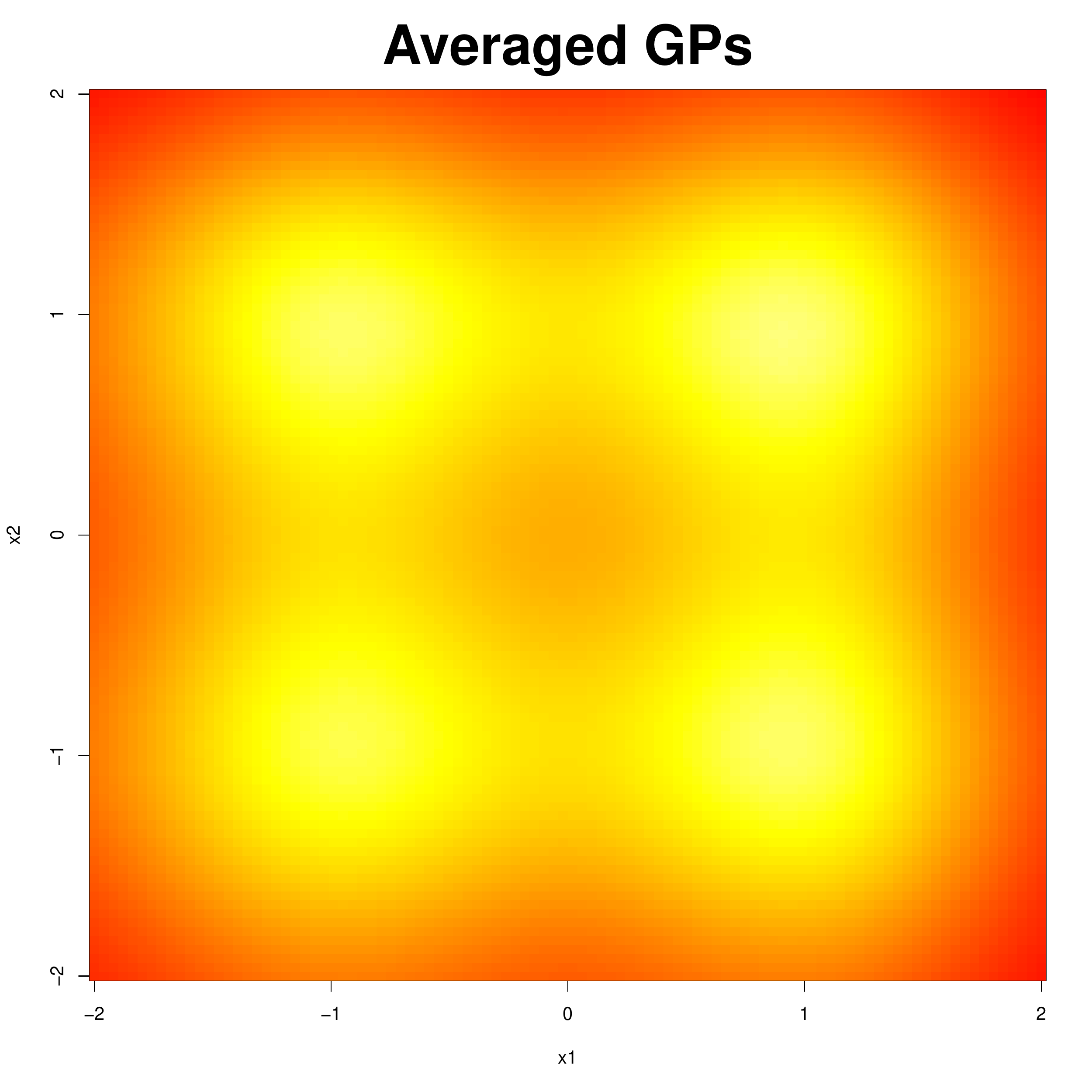}%
		\includegraphics[width = .5\textwidth]{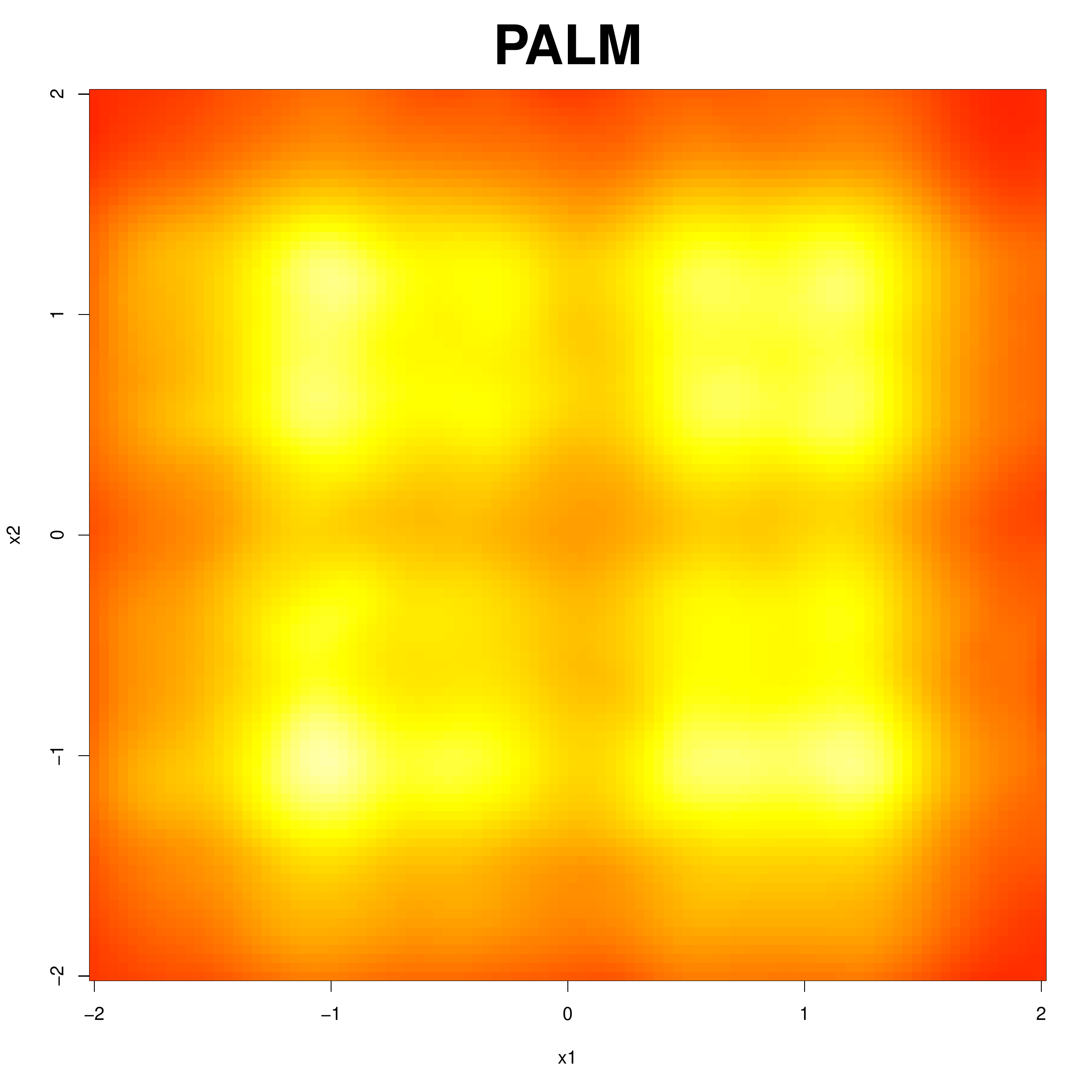}
		\vspace{-0.75cm} \caption{Model averaging (left) and PALM
		(right) predictions on Herbie's tooth.}
	\label{fig:schemes} \end{figure}

	Recombining component models into a unified prediction can work even if the
	data are not partitioned, as long as independence of the model, given a
	global function, can be maintained [\cite{chen_bagging_2009}]. The
	computational complexity for fitting and storage of the model are reduced
	like in a typical partition model, while prediction is $K$-fold more work
	because every predictor is involved for each predictive location $x$.  A
	drawback, however, is that a common method for justifying functional
	independence is to have the training sets in each component being quite
	sparse relative to the global training data.  This limits the weighted
	average's ability to accurately capture complicated trends, and results in
	over-smoothing.  To illustrate, consider the left panel of Figure
	\ref{fig:schemes} which is based on precision averaged global models. The ten
	thousand point training set is randomly partitioned into 100 models
	containing 100 points each.  Each model's prediction is weighted by a
	softmax on precision.  Compared to either the left panel of Figure
	\ref{fig:error} via LAGP, or Figure \ref{fig:part} via partitioning,
	fidelity is clearly lost.  Smoothly varying (e.g., precision) weights
	guarantee continuity of the surface, but emphasis on local ``authority of
	influence'' is starkly absent.

	Therein lies the setup for our methodological contribution: PALM
	(precision aggregated local model). The idea behind PALM is to build
	locally focused models, as in a partition scheme or LAGP, but which can be
	smoothly recombined into a global model.  The framework introduced in
	Section \ref{sec:palm} momentarily could apply in principle with any local
	model furnishing smooth predictive means and variances, but our
	presentation and empirical work favor LAGP.  Our goal is to retain the
	accuracy of an LAGP predictor, but lean on the PALM framework to obtain
	smoothness and computational gains.   Appropriate weighting and accounting
	for covariance is highly component-model specific and we provide details
	for LAGP, however similar analogues will be readily apparent.  The right
	panel of Figure \ref{fig:schemes} serves to whet the appetite.  PALM
	offers a hybrid between predictors shown in earlier figures.  Focusing on
	the slice in the right panel of Figure \ref{fig:error}, observe that bias
	in these out-of-sample residuals is lower, smoother, and has less
	recognizable pattern than under LAGP.


	\section{PALM methods} \label{sec:palm}

	In its most general form, PALM is
	a method of dynamically weighting predictions from a series of locally
	accurate models, or {\em local experts}, to create a coherent global model.
	Our emphasis is on GP local experts, spaced evenly throughout the input
	space, but we see no reason why a similar scheme may not be applied more
	widely. Eq.~(\ref{eq:partvar}) provides the basic
	framework; PALM addresses how to choose weights, $w_k(x)$, in order to create
	smooth and accurate predictions, as well as methods to estimate covariance
	terms: $\sigma_{kj}(x)$, for $k \neq j$.  It is important to reiterate that
	partitioning schemes (Section \ref{sec:partagg}) are a special case of PALM
	where one selects $w_k(x) = 1$ for exactly one of $k=1,\dots,K$, for each
	$x$, and zero otherwise.  That weighting eliminates the need to estimate
	covariance terms $\sigma_{kj}(x)$.  On the flip size, a partition model can
	be converted into a smooth PALM by swapping in smoothly varying weights and
	estimating between model covariances.

	\subsection{Warmup: surrogate modeling deterministic evaluations}

\begin{figure}[ht!]
		\centering
		\includegraphics[width=.34\textwidth,trim=10 10 25 0]{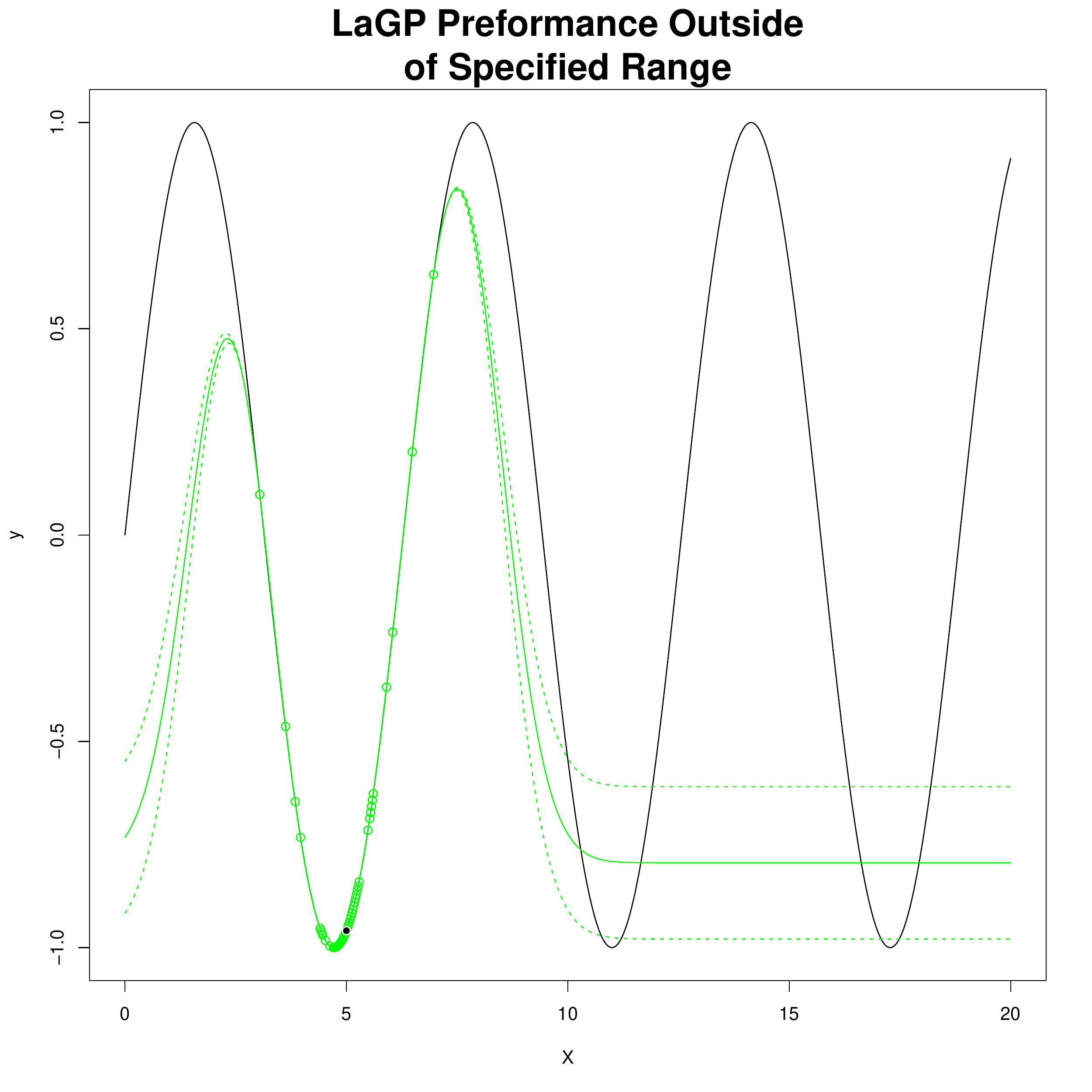}%
		\includegraphics[width=.32\textwidth,trim=50 10 25 0,clip=TRUE]{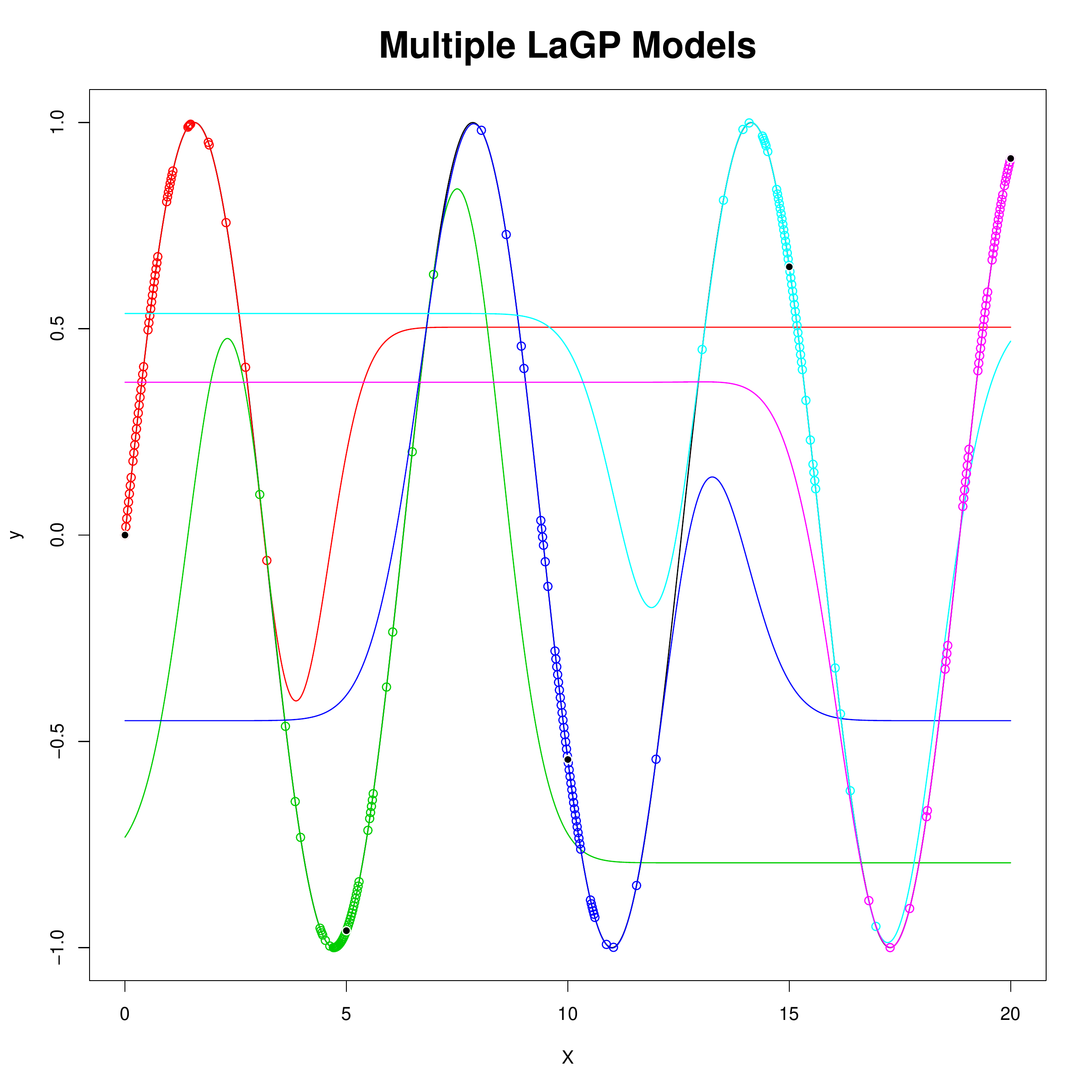}%
		\includegraphics[width=.32\textwidth,trim=50 10 25 0,clip=TRUE]{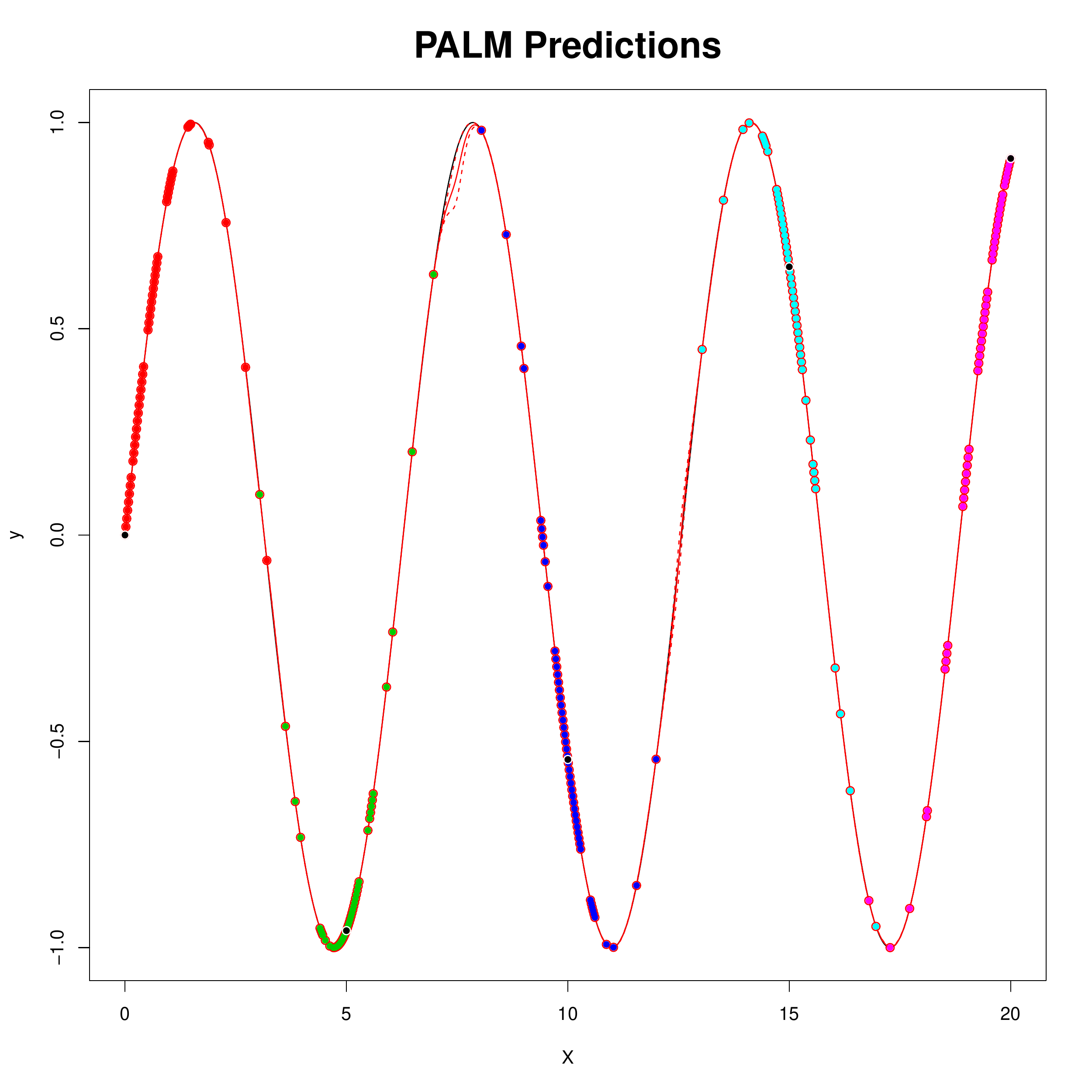}
		\vspace{-0.25cm}
		\caption{A sine wave modeled by LAGP predictors: alone at
		one location (left); at five locations (middle); after PALM
		weighted averaging (right). } 
		\label{f:sin}
	\end{figure}

	Local GP approximation, via LAGP, is designed to predict at a specific point
	$x$, but in fact that fit offers high quality predictions over a much
	broader area of the input space.  To illustrate, consider the left panel
	Figure \ref{f:sin} which shows how an LAGP model can offer effective
	surrogate modeling (prediction and uncertainty quantification) of a sine
	wave $x \in (3,7)$ despite being trained only at the ``center'' value of
	$x_k = 5$. The objective function is comprised of 1000 equally-spaced points
	from a sine wave on $[0,20]$.  The training subdesign is derived from
	default settings in the {\tt laGP} package: starting with six nearest
	neighbor (NN) points, the ALC criteria is used to greedily select subsequent
	points before reaching a final size of 50. In the middle panel, the same
	idea is applied at four other centers spaced evenly throughout the input
	space. Finally, in the panel on the right, these $\mu_k(x)$ and
	$\sigma_k^2(x)$, for $k=1,\dots,K=5$, are combined with inverse variance
	weights (\ref{eq:partvar}), details coming momentarily,
	to yield a global predictor which is quite accurate throughout the domain of
	interest despite a non-uniform density of sampling. The overlapping regions
	of these five local experts is enough to cover prediction for the entire
	region.

	A one dimensional example offers convenient visuals --  overlapping
	predictive areas of multiple LAGP and their PALM combination together with
	an underlying truth -- but the same principles hold true for higher
	dimensions, although visualization is more challenging.  The right panel
	of Figure \ref{fig:schemes} shows a PALM fit to Herbie's tooth
	(\ref{eq:herbie}). The training set is based on a 5000 element regular
	grid, and the ``centers'' for 100 LAGP local experts (again via {\tt laGP}
	defaults) were chosen via maximin design\footnote{The maximin criteria was
	extended to include distance to the boundary in order to build a buffer to
	prevent centers of {\tt laGP} models on the edges of the space, which
	would effectively clip their area of expertise.}
	\citep{johnson1990minimax}. Those local designs are not shown in the
	figure, because there would have been too much clutter.  However, some are
	shown later in Section \ref{sec:weights} when discussing details of the
	PALM weighting scheme.  The testing set was a regular grid of size ten
	thousand. Full LAGP prediction on those locations takes twelve (core)
	minutes; our PALM analog takes six seconds.

	The right panel of Figure \ref{fig:error} shows a slice through this
	surface, comparing error from LAGPs trained to every element of that slice
	of the predictive grid.  Notice how the PALM predictor is both more
	accurate, and also exhibits a smoother bias/higher accuracy compared to the
	truth. Section \ref{sec:empirical} presents more examples, but first it
	makes sense to deliver more detail on how PALM patches together a limited
	number of local experts to obtain smooth, more accurate prediction, than
	local experts trained exhaustively in the predictive grid.

	\subsection{Choosing weights} \label{sec:weights}

	Choosing a smooth weighting function that simultaneously selects and
	hybridizes between the right local experts, while maintaining continuity in
	both mean and variance surfaces, is key to the PALM framework.  Inverse
	variance weighting $w_k(x) = \phi_k(x)/\sum_{\ell=1}^K \phi_\ell (x)$,
	introduced in Section \ref{sec:avg}, offers a good starting point, combining
	smoothness and localization according to the local experts' own judgment.
	Locally trained GP predictors, e.g., from LAGP, offer an organic and smooth
	reduction in variance nearby to the local design they are trained upon, and
	consequently will receive higher weight there.  Conversely, higher variance
	away from their area of expertise will result in lower weight if another
	expert is commensurately better there.  A downside, however, is that GP
	models are not unbiased -- they revert to the prior process, and in
	particular the {\tt laGP} implementation uses a prior mean of zero.  Such
	bias, then, would become more severe as local GP experts extrapolate.

	This means that theory of optimality of inverse-variance weights
	\citep{cochran_combination_1954} does not apply, but coupled with that
	is perhaps a more practical issue: an inevitable compression of weight
	in regions of the input space where disparate local experts are equally
	good/bad.  Intuitively, a weight $\phi_k(x)$ should increase nearby a
	simultaneously shrinking domain of expertise around $x$, i.e., have
	weight decreasing elsewhere.  However, this cannot happen with a GP
	expert because predictive variance is also prior reverting.
	Consequently, adding local experts into PALM, no matter where their
	local domain of expertise lies, has a flattening or ``cooling'' effect
	on individual weights.  To combat both issues (bias and cooling) we
	propose raising precisions to a power $p$ before applying the softmax,
	i.e., $w_k(x) = \phi_k(x)^p/\sum_{\ell=1}^K \phi_\ell(x)^p$, where $p$
	is a function of the number of local experts, $K$, and the size of the
	input space, i.e., it's dimension $d$ assuming inputs coded to
	$[0,1]^d$.  In particular, we suggest $p = \log_d K$ as a sensible
	automatic choice, however another option is to treat $p$ as a tuning
	parameter which could be optimized on a hold out set.  

	\begin{figure}[ht!]
		\centering
		\includegraphics[width=.4\textwidth]{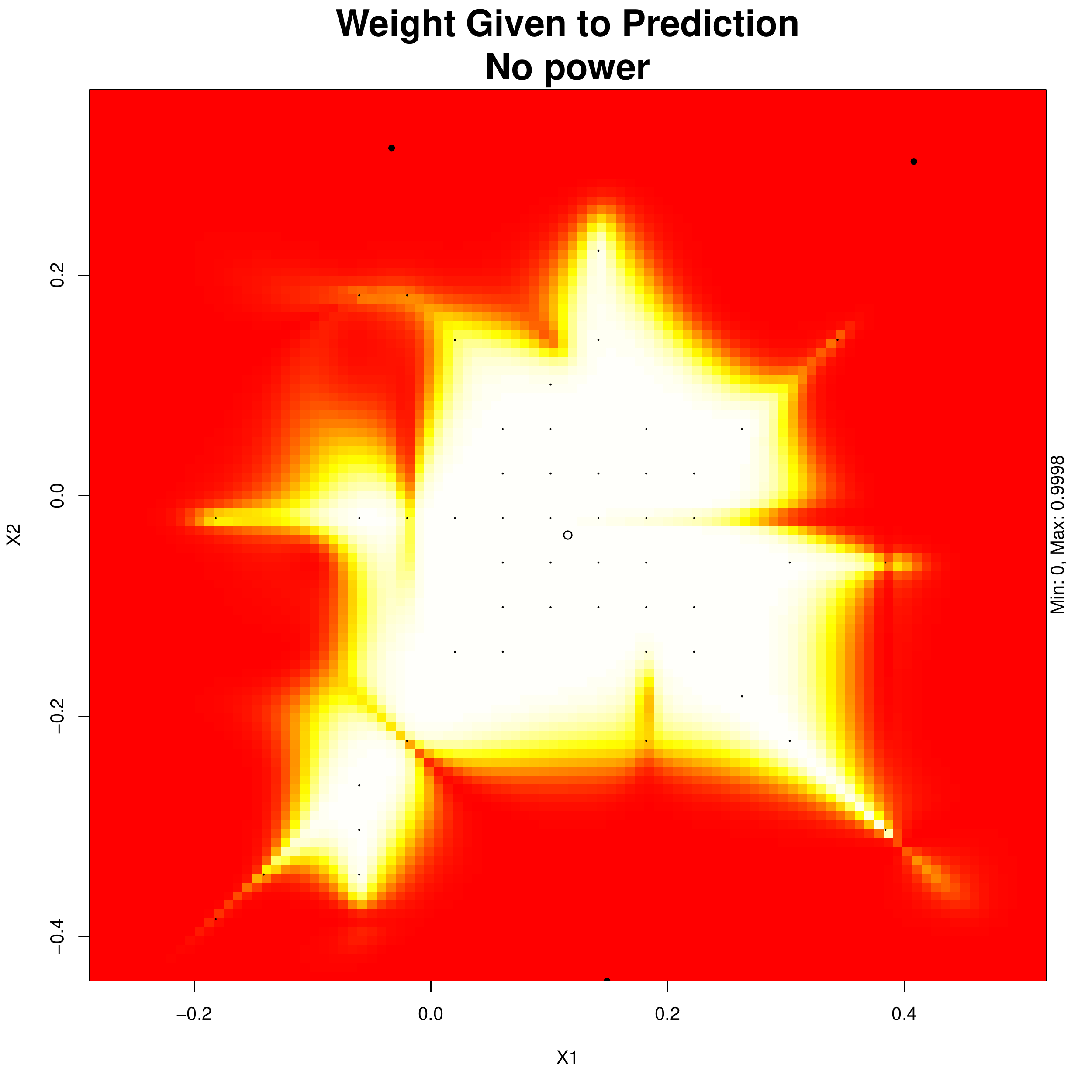}%
		\includegraphics[width=.4\textwidth]{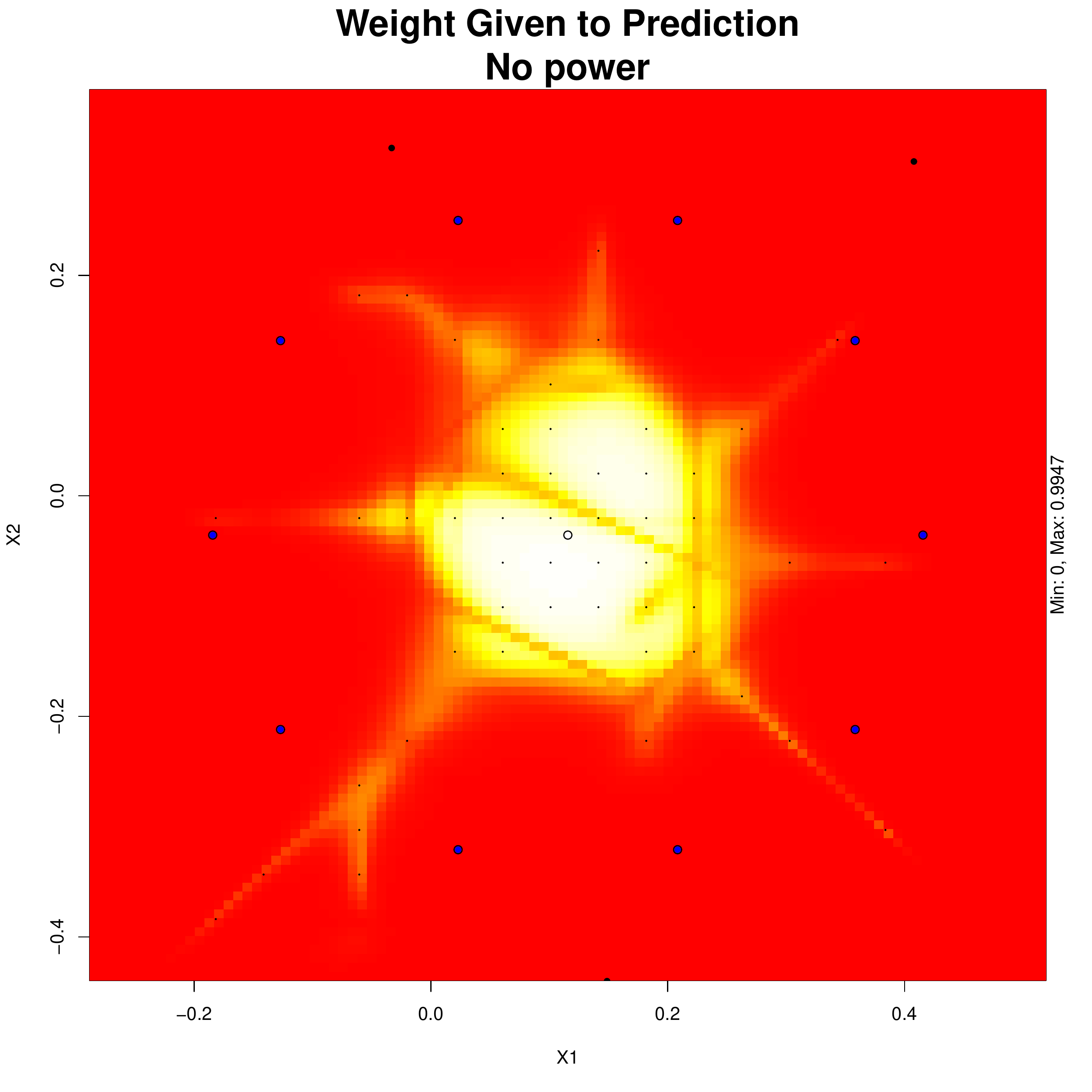}\\
		\includegraphics[width=.4\textwidth]{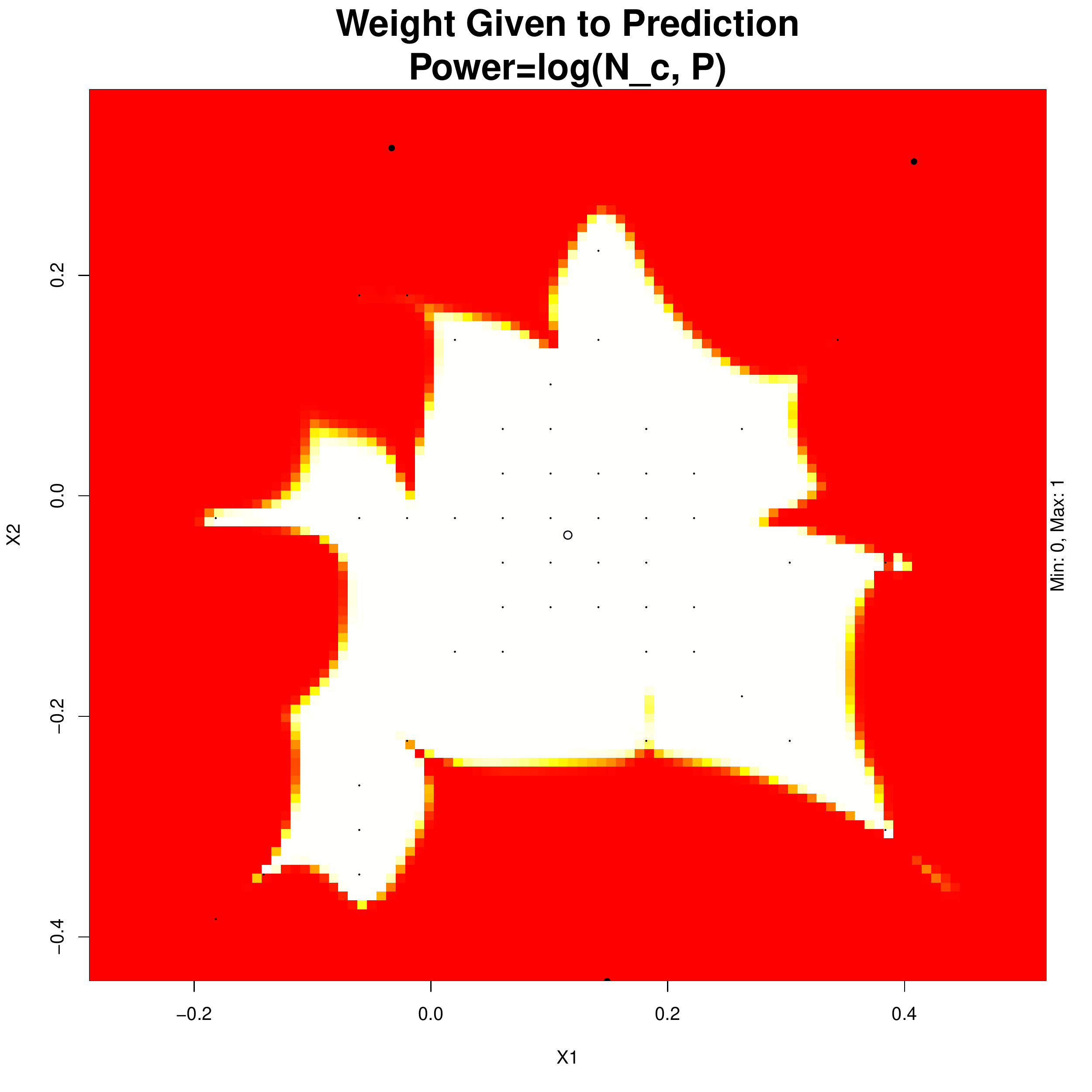}%
		\includegraphics[width=.4\textwidth]{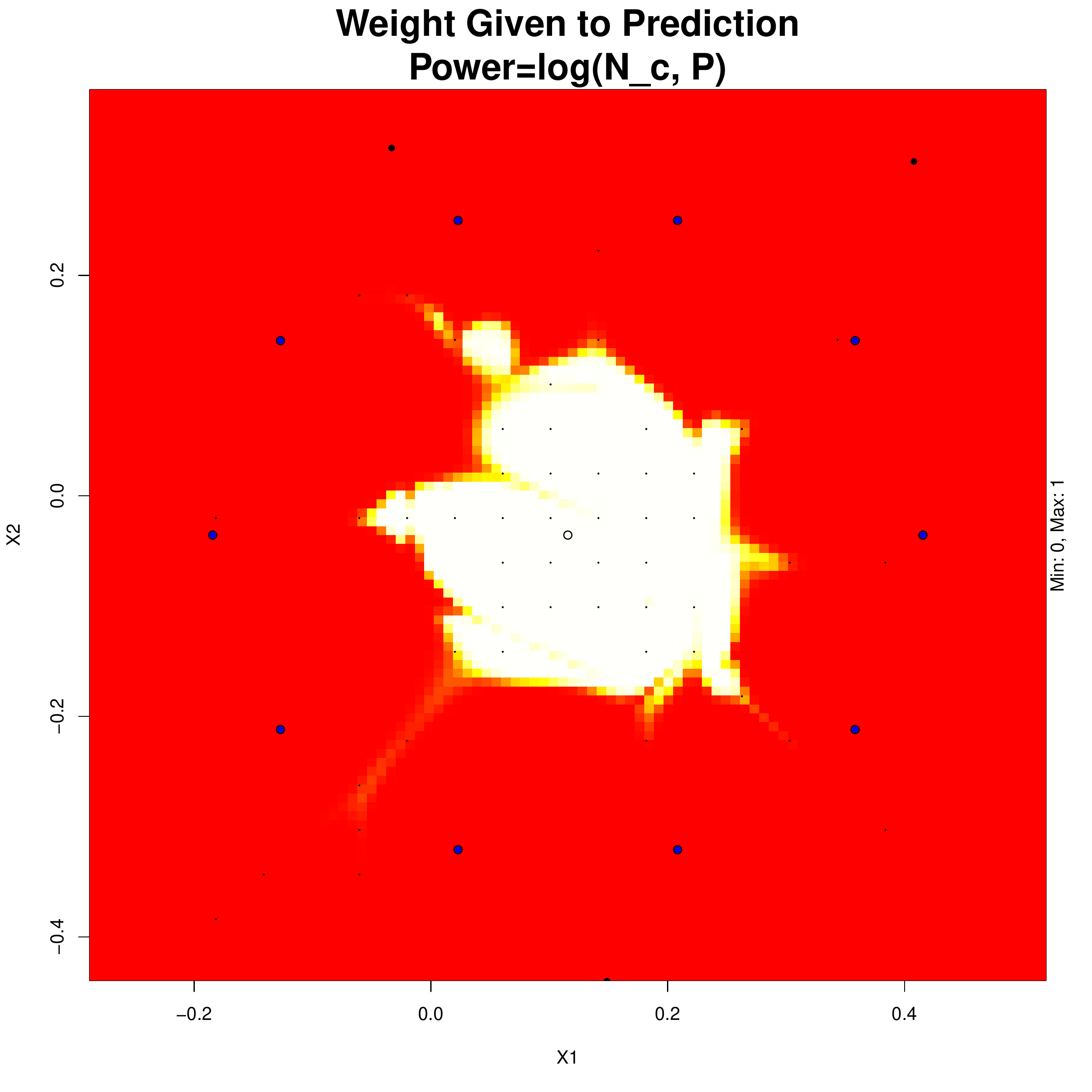}
		\vspace{-0.25cm}
		\caption{Top left: Weight given to a local expert with $p=1$,
		i.e., un-powered.  Top right: decrease in weight caused
		by adding extra models outside of the area of expertise.
		Bottom left: Local expert weights when powers
		are used.  Bottom right: Added centers outside of the area of
		expertise do not affect the weight.  Large black/blue dots
		represent other local experts in the PALM, while small dots
		represent the training inputs $X_N$.}
		\centering
		\label{f:weight}
	\end{figure}

	Figure \ref{f:weight} illustrates the effect of powering up weights. What we
	see in all panels are the predictive weights given to one LAGP expert's
	``center'' (open circle) of the PALM used in Figure
	\ref{fig:schemes}.  The figure has been zoomed into the local expert, and
	the centers of other local experts used in the original model are
	represented by commensurately sized filled circles. Smaller dots nearby the
	open circle indicate the local design selected by LAGP for the open-circle
	expert. Red colors in the image plot(s) indicate low weight, and
	lighter/whiter colors indicate high weight.  Now consider each panel
	individually.  The top left panel corresponds to softmax precision weights,
	i.e., $p=1$, not powering up.  For contrast, the bottom left panel uses our
	recommended power instead, $p = \log_d K$. Observe how powering up has a
	``heating'' effect, making weights more extreme on both ends, leading to
	fewer yellow/orange and more white/red.  In particular, weight is greatly
	increased in the region nearby the central expert. Otherwise the same of the
	area with non-trivial (red) weight is largely unchanged.

	Now consider the panels on the right in the figure which consider the effect
	of adding new local experts into the PALM fit.  The ``centers'' of those
	experts are indicated by larger blue-filled dots. The top right panel
	corresponds to no powering up ($p=1$), and the bottom right one uses $p=
	\log_d K$. Otherwise everything is the same as on the left; weights
	correspond to the central expert indicated by the open circle, etc.  In the
	top right, notice that even though the new experts are added largely outside
	of the original ``center's'' area of expertise, the weight for the central
	expert has gone down, in relative terms (i.e., after re-normalization)
	everywhere.  The non-red area is smaller, and the proportion of yellow is
	much larger than in any other panel in the figure. Not only has weight gone
	down within that expert's domain of expertise but, intriguingly, the
	presence of the new experts have created streaks of low weight in regions
	that are very close to those new experts.  (The new experts are somehow not
	experts nearby themselves, albeit in a small volume.)   This indicates a
	lack of ability for pure precision weights to accurately determine which of
	our experts should be trusted.  In the bottom right panel, after
	appropriately powering up precisions, we see that the same ring of new
	experts has lowered the weight only inside their own areas of expertise,
	whereas weight given to the central model remains high in the densely packed
	region of design points. More experts mean more reactivity in domain of
	expertise across the input space.
	
	\begin{figure}[ht!]
		\centering
		\includegraphics[width=.45\textwidth]{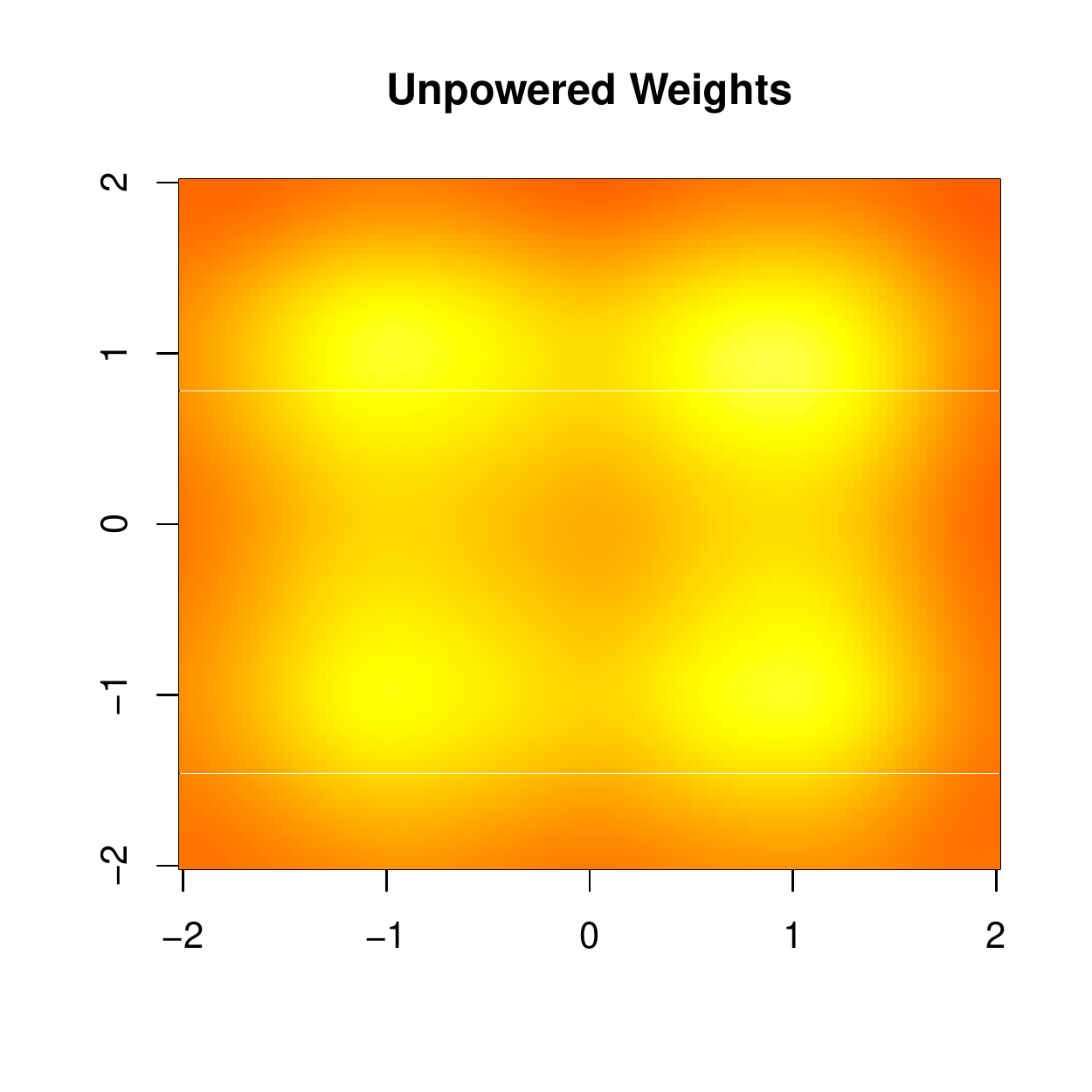}%
		\includegraphics[width=.45\textwidth]{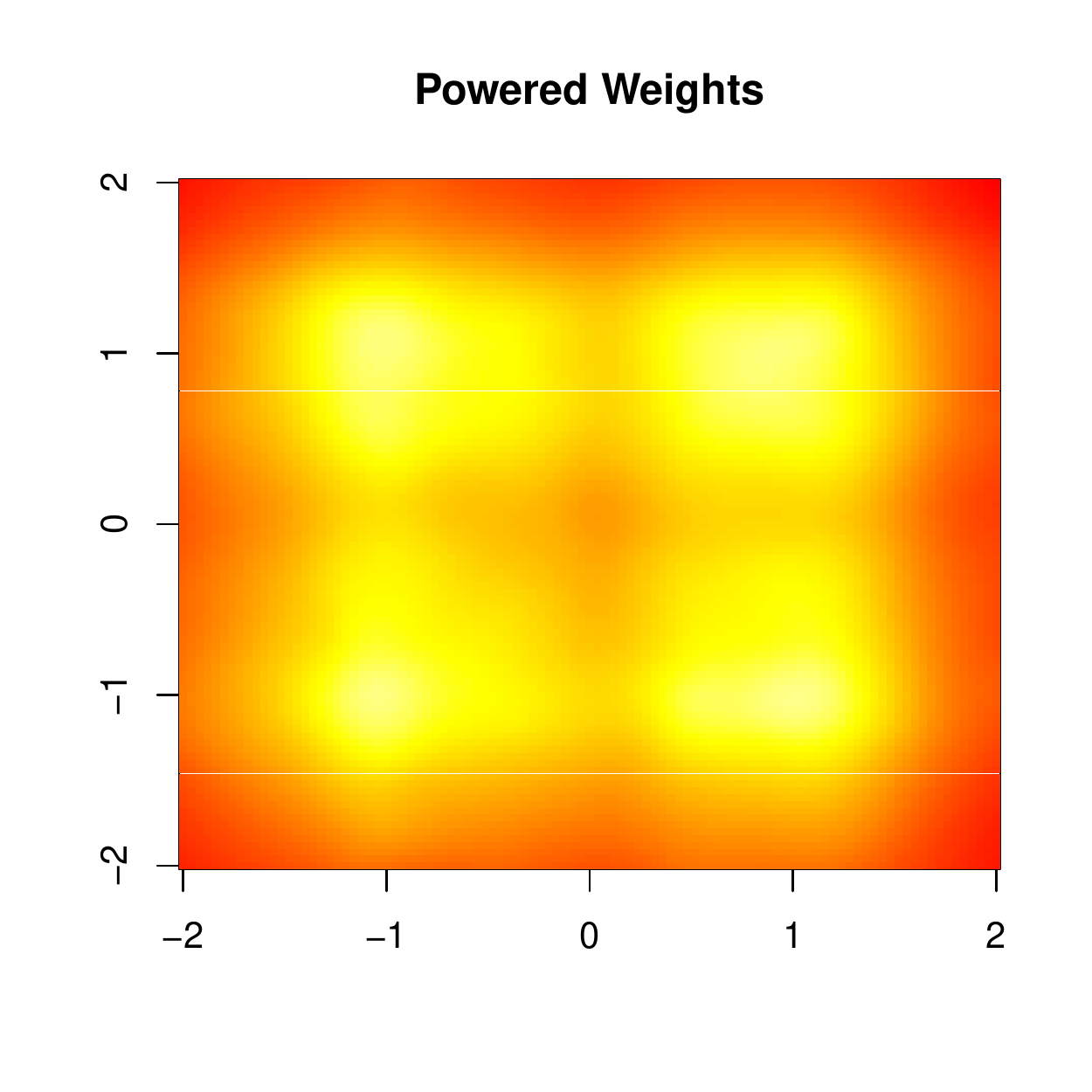}
		\vspace{-0.75cm}
		\caption{Left: Predictions from a palm using $p=1$, i.e. pure precision weights.  Right: Predictions from the same PALM using $p=\log_2 K$.  Local model predictions are unchanged.  Only the weight is different.}
		\centering
		\label{f:weight2}
	\end{figure}
	
	The effect this has on model prediction is non-trivial. Observe the left
	panel of Figure \ref{f:weight2}.  Here we see PALM predictions on Herbie's
	tooth (following the given experimental details) with pure precision
	weights, i.e. $p=1$.  Although the rest of the global PALM predictor is
	performed to the exact specifications laid out here, the model exhibits
	the same global cooling behavior as the averaged predictor shown above.
	In the right panel, we apply our recommended $p=\log_d K$ weights,
	resulting in vastly improved predictive performance.

	\subsection{Estimating covariance} \label{sec:cov}

	A major difference between PALM and other model averaging schemes is that
	we cannot assume functional independence.  Other schemes get away with
	approximate functional independence because each expert $k$ is learning
	something broad, $\hat{f}_k$, which is a good approximation to the global
	function, $f$.  Consequently, globally focused averaging schemes have a
	relatively constant covariance between models across the input space that
	authors claim to be low, and tacitly approximate as zero without
	noticeably ill effect. Locally focused models, on the other hand, can have
	an extremely high covariance in areas where their domains of expertise
	overlap, and low elsewhere.  An illustration in the LAGP case is coming in
	Figure \ref{fig:corr}, a little later in Section \ref{sec:details}.
	Because of this, we need a way to estimate the covariance between local
	experts. Assuming zero covariance will result in underestimating the
	uncertainty in our predictions (\ref{eq:partvar}).  A challenge in
	estimating those covariances is that our local experts assume independence
	when performing local design and inference.  Therefore observing
	co-variability must transpire as a post-processing step.

	Toward estimating those covariances, we begin by decomposing covariance into
	the product of standard deviations and correlation: $\sigma_{kj}(x) =
	\sigma_k(x) \sigma_j (x) \rho_{kj}(x)$. The two standard deviations,
	$\sigma_k(x)$ and $\sigma_j(x)$ are furnished by the predictor from each
	local expert, so we only need a way to additionally estimate the
	correlation, $\rho_{kj}$.  Pointwise, for each $x$ and experts $j$ and $k$,
	we desire an estimate of the correlation between every pair of models:
	\[ 
	\rho_{kj}(x) =
	\mathbb{C}\mathrm{orr}(\hat{\mu}_k(x), \hat{\mu}_j(x)) =
	\frac{\mathbb{E}\{(\hat{\mu}_k(x) - Y(x))(\hat{\mu}_j(x) -
	Y(x))\}}{\hat{\sigma}_k(x) \hat{\sigma}_j(x)}.
	\] 
	For many models, $\rho_{kj}(x)$ may be difficult to estimate, especially
	when ``lazy evaluation'' at $x$ is required, i.e., when we don't yet know
	where we wish to predict at fitting time.  One of the main aims in PALM
	front-load calculations at training and avoid such expensive calculation at
	predict time.  One way is by estimating a single $\hat{\rho}_{kj}$ for all
	$x$, and for all $j,k = 1,\dots,K$, imposing a form of stationarity in
	correlation over the testing set.  Because correlation between models is not
	constant over the input space, and is only high in the areas where a pair of
	models have overlapping influence, the non-overlapping space on the set
	$X_{kj} = X_k \cup X_j$ will artificially drag correlation down.
	Combined with powered up weights, which helps to ensure that two models
	will only receive high weight in the area where their influence overlaps,
	this empirical estimation compounds into a poor estimation of the
	correlation in exactly the area where the value has the most impact on the
	model. Again we delay a specific example until Section \ref{sec:details}.

	We find that a suitable stationary value to plug in for $\hat{\rho}_{kj}$ is
	the maximum value over the input space: $\hat{\rho}_{kj} = \max_{x}
	\rho_{kj}(x)$. This formulation, however, becomes increasingly hard to
	estimate in higher dimension.  Moreover $\rho_{kj}(x)$ depends on knowledge
	of $Y(x)$, a chicken-or-egg-problem at the time of fitting.  Thus the final
	layer of approximation is to replace continuous maximization $\max_x$ with
	discrete search over data $X_{kj}$ used to train local models $j,k$.  That
	is, given a class of models for which we can obtain a point-wise estimate of
	model correlation, $\hat{\rho}_{kj} = \mathbb{C}\mathrm{orr}(\hat{\mu}_k(x),
	\hat{\mu}_j(x))$, we use: 
	\[ 
	\hat{\rho}_{kj} = \max_{x \in
	X_{kj}}\left\{\rho(\hat{\mu}_k(x), \hat{\mu}_j(x))\right\}.
	\] 
	Using this formulation, we obtain a high estimate of correlation if either
	model contains a point in the overlapping area, and a low estimate
	otherwise.  Specific forms for $\rho(\hat{\mu}_k(x), \hat{\mu}_j(x))$ depend
	the local expert's predictive equations; details are given for LAGP
	in Section \ref{sec:details}.

	We recognize this is an approximation, but our weighting scheme ensures
	than any inaccuracies would have a minor impact on the uncertainty
	captured by the PALM predictor.  In Eq.~(\ref{eq:partvar}), observe
	that the impact on $\mathbb{V}\mathrm{ar}(x)$, for a single pairing of
	$k \neq j$, follows
	\begin{align*}
		w_k(x) w_j(x) \hat{\sigma}_k(x) \hat{\sigma}_j(x)
		\hat{\rho}_{kj} &\propto \left(\hat{\sigma}_k^{-2}\right(x))^p
		\left(\hat{\sigma}_j^{-2}(x)\right)^p \hat{\sigma}_k(x)
		\hat{\sigma}_j(x) \hat{\rho}_{kj} \\
		&= \hat{\sigma}_k^{-2p+1}(x) \hat{\sigma}_j^{-2p+1}(x)
		\hat{\rho}_{kj}.
	\end{align*}
  If we pick a form for $\hat{\rho}_{kj}$ which is not too small,
	i.e., $\hat{\rho}_{kj} \geq \hat{\rho}_{kj}(x)$, we get sensible
	results from the perspective of conservative
	estimates of uncertainty.  Examine the case where two local models have
	an overlapping area of extremely high correlation, i.e.,
	$\max_{x}(\rho_{kj}(x)) \approx 1$, which is the worst case scenario
	for the disparity between our stationary estimate and reality.  When we
	predict close to the area where $\rho_{kj}(x) = \max(\rho_{kj}(x))$,
	the true correlation will be close to the maximum and the value we have
	chosen will benefit the model.  As we move away from this area,
	$\left(\sigma_k \sigma_j\right)^{-2p+1}$ decreases faster than the
	disparity between that max and the true correlation.  This squashes the
	piece of the model variance that is due to this covariance.

	\subsection{Implementation details} \label{sec:details}

	The PALM introduction aimed generic, or agnostic to the choice of expert
	although LAGP featured as an exemplar.  Fixating specifically on that
	choice, several details must be in place in order to fully operationalize
	the methodology.  Fitting a GP model, local or otherwise, requires selecting
	hyperparameter settings, such as lengthscales $\theta^k = (\theta^{(k)}_1,
	\dots, \theta^{(k)}_d)$ and scale $\tau_k^2$.  The {\tt laGP} library
	returns fitted values for these, independently for each predictive location
	or ``center'' for PALM.  These are designed for predicting at one site only,
	without knowledge of how PALM intends to utilize those predictors more
	widely. It may be sensible to revise those estimates in light of the global
	scope of the wider PALM predictor.  Local variation in lengthscales
	$\theta_k$ is sensible, and we have found no need to make adjustments for
	the PALM setting, except to tailor {\tt laGP}'s prior to discourage
	extremely large lengthscales.  Assuming we have no knowledge of the global
	lengthscale, we can find a reasonable estimate by building several global GP
	models using small random subsets of the data, and selecting the largest
	maximum likelihood estimate from among them to be the maximum value that
	each {\tt laGP} may choose.  Local experts may then adjust downward as
	necessary.

	Scale $\tau^2_k$ requires a more careful treatment, however, since it
	determines the amplitude of variability in the predictor (\ref{eq:mux}).
	When trained on a local design, $\hat{\tau}_k^2$ could only reflect the
	scale of the local training design.  When applied globally through
	predictive equations, this could be a mismatch to a much broader global
	scale, and thus result in extreme inaccuracy especially in terms of
	predictive variance. This is particularly important since predictive
	variance is used to determine the weight of local models upon recombination
	through the softmax.  Therefore, instead of estimating a separate amplitude
	for every model, we prefer to leverage information from all local experts to
	enforce a single hyperparameter value.  In other words, we are building in
	an assumption of global stationarity in scale. One option here
	which is attractive theoretically is to perform inference for a global
	$\tau^2$ by maximum likelihood, where the likelihood is comprised of a
	product of $K$ MVN densities whose means and covariances follow $K$
	applications of Eq.~(\ref{eq:mux}), one for each local expert.  In practice,
	however, this is intractable for even modestly sized training sets $N$,
	owing to the requisite cubic decomposition of $N \times N$ matrices.
	Instead, we prefer the following idea.

	Hyperparameter $\tau^2$ doubles as amplitude and asymptotic variance
	$\tau^2(1 + \eta)$ of a GP model, measured as the correlation to training
	data $X_N$ decays to zero. For now, take nugget as jitter for interpolation
	of simulations observed without noise, $\eta =
	\varepsilon$, however details for the noisy case will be provided
	momentarily. Reverse engineering a bit then, consider measuring that
	asymptotic variance through the PALM predictor instead.  Below $s^2$
	represents the empirical variance of the response vector, $y$, via
	simple residual sum of squares around the average $\bar{y}$.  The plan
	is to use that estimate to define what the asymptotic variance of the
	final PALM fit should be. Let quantity $\tau^2_{\mathrm{PALM}}$
	represent the effective amplitude parameter of the full PALM, while
	$\tau^2$ is the amplitude we use to fit the local experts in order to
	achieve the correct extrapolation properties.

	What we want is for the asymptotic amplitude of our PALM fit to be
	equal to the measured amplitude of the training data.  In other words,
	we want $\tau^2_{\mathrm{PALM}}(1+\eta) = s^2(1+\eta)$.  In order to
	achieve this, begin with the formula for variance from the PALM fit
	(\ref{eq:partvar}) \begin{align*} \hat{\sigma}^2(x) &= \sum_{k = 1}^K
	\sum_{j=1}^K w_k(x) w_j(x) \hat{\rho}_{kj} \hat{\sigma}_k(x)
	\hat{\sigma}_j(x), \end{align*} with covariance estimates (correlation
	and standard deviations) plugged in. It is important to remark that the
	weight function, $w_k(x)$ depends on the GP correlation function
	between the data used to train model $k$, and the predictive location
	$x$ as $k_k\left(X_k, x\right)$ through the GP variance function,
	$\hat{\sigma}_k^2(x) = \tau^2 (1 + \eta - k_k^\top(x) K_k^{-1} k_k
	(x))$.  As predictive location $x$ moves away from the corpus of data
	used to train PALM experts (a subset of the full training data), each
	of the GP correlation functions approach 0.  This causes the weight
	each PALM ``center'' receives as they extrapolate to be approximately
	equal, i.e., uniform.  Consequently, we may replace those weight
	functions in the double sum above with $1/K$ where $K$ is the total
	number of experts in the full PALM. 
	\begin{align*}
		\mbox{As } \quad
		\lim\limits_{k\left(X_k, x\right) \rightarrow 0} w_k(X) =
		\frac{1}{K}.
		\quad \mbox{ we have that } \quad \hat{\sigma}^2(x)
		&\rightarrow \frac{1}{K^2}\sum_{k = 1}^{K}\sum_{j=1}^{K}\rho_{kj}
		\hat{\sigma}_k(x) \hat{\sigma}_j(x).
	\end{align*}
	We may further use the limit of the GP correlation function to remove
	$\hat{\sigma}_k(x)$ from the double sum.  As each $k_k^\top(x) K_k^{-1}
	k_k (x)$ approaches 0, $\hat{\sigma}_k^2(x) \rightarrow \tau^2 (1 +
	\eta)$, simplifying to 
	\begin{align*}
		\hat{\sigma}^2(x) \rightarrow
		\tau^2\left(1+\eta\right)\frac{1}{K^2}\sum_{k =
		1}^{K}\sum_{j=1}^{K}\rho_{kj}.
	\end{align*}

	Finally, remembering that we want to pin this value to a reasonable
	amplitude based on the variance of the observed response vector, we can
	solve for $\tau^2$, the amplitude that should be plugged into each
	individual local expert.  Specifically, 
	\begin{align*}
		s^2(1+\eta) &= \tau^2\left(1+\eta\right)\frac{1}{K^2}\sum_{k =
		1}^{K}\sum_{j=1}^{K}\rho_{kj}, &
		\mbox{giving } \quad	\tau^2 &= s^2
		\frac{K^2}{\sum\sum\rho_{kj}}.
	\end{align*}
	The final ingredient above, circling back the generic discussion in Section
	\ref{sec:cov}, is estimating $\rho_{kj}$, the correlation between LAGP
	experts. Define the {\em predictive kernel} of each expert $j$ to be the
	matrix wise correlation between each point in the model, and predictive
	location $x$:
	\begin{equation}
		k_j(X_j, x)^\top K_j k_j\left(X_k, x\right).
		\label{eq:gppredcor}
	\end{equation}
	For each pair of models $kj$, we may find the kernel
	when using one model to predict the other, and visa versa.  A sensible
	estimate for between-expert empirical correlation is the maximum value
	found in the combined vector of predictive kernels.  Let
	$\rho_{kj}^\ell = k_k\left(x_\ell \right) K_k^{-1} k_k\left(x_\ell
	\right)$, where $\ell$ indexes each $x_\ell \in X_j$. Then, take
	\begin{equation}
		\hat{\rho}_{kj} = \max\left(\rho_{kj}^1, \rho_{kj}^2, \dots,
		\rho_{ij}^{|X_j|}, \rho_{jk}^1, \rho_{jk}^2, \dots,
		\rho_{jk}^{|X_j|} \right). 
		\label{eq:rhohat}
	\end{equation}

\begin{figure}[ht!] \centering
		\includegraphics[width=.7\textwidth,trim=0 45 0 0,clip=TRUE]{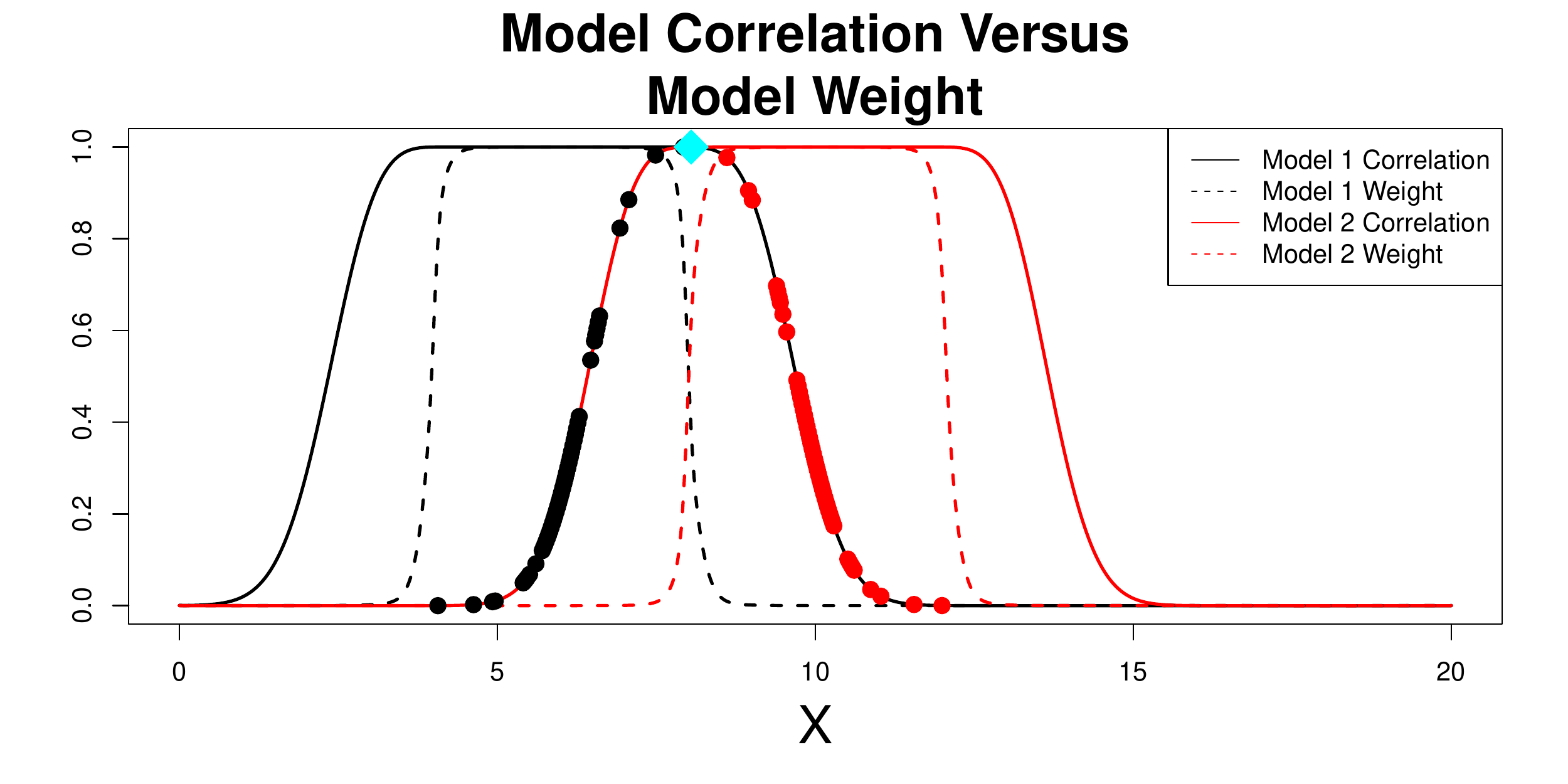}
		\caption{The correlation chosen between two local experts
		overlaid with the weight they receive in the model.  The estimated
		correlation is shown as a blue diamond for
		visibility.}
		\centering
		\label{fig:corr}
	\end{figure}

	This kernel based correlation estimate (\ref{eq:gppredcor}) is bounded on
	$[0,1]$, and is highly positive for models that are close to each other,
	while approaching zero for those that are separated.  To illustrate this,
	Figure \ref{fig:corr} displays two local experts used to model the sine
	wave from Figure \ref{f:sin}.  Note that these are not the same models
	from above, but have been moved closer together to ensure they overlap for
	dramatic effect.  Solid lines represent the GP correlation
	(\ref{eq:gppredcor}) between each model and every point along the wave.
	Dashed lines show the weight that each model receives in predicting along
	the function.  We have put the points used to build local experts along
	the correlation line for the opposite model, and colored the maximum point
	(which is used as the value for $\hat{\rho}_{kj}$) in light blue.  Notice
	the value we choose for $\hat{\rho}_{jk}$ following Eq.~\ref{eq:rhohat},
	shown as a blue diamond for easy visibility, is very close to 1, which is
	accurate for the area of weight overlap.  This value may not be accurate
	for the entire input space, but one of the models receives nearly zero weight
	for all egregious areas.
	Using estimates $\hat{\rho}_{kj}$ we calculate
	predictive variance at any new point using 
	\[ 
	\hat{\sigma}^2(x) =
	(\vec{w}(x)*\vec{\sigma}(x))^\top \hat{\rho}
	(\vec{w}(x)*\vec{\sigma}(x)) 
	\] 
	where $\vec{w}(x) = (w_1(x), w_2(x),
	\dots, w_K(x)\}$, $\vec{\sigma}(x) = \{\hat{\sigma}_1(x),
	\hat{\sigma}_2(x), \dots, \hat{\sigma}_K(x)\}$, ``$*$'' is a
	component wise, Hadamard product, and $\hat{\rho}$ is the matrix
	containing empirical correlation values.

	\subsubsection*{Regression for noisy data}

	The major difference between a deterministic GP model and the stochastic
	version is the incorporation of a nontrivial nugget value, $\eta$, which
	breaks interpolation for smoothing, allowing for variance at training data
	points.  The {\tt laGP} function can return an estimated nugget for each
	local expert, obtained by maximum likelihood.  This ignores that we intend
	to use the this local expert as part of a unified model.  It's also overkill
	because separate nuggets mean heteroskeasticity \citep[i.e., input-dependent
	noise][]{goldberg:williams:bishop:1998,binois2018practical}, and thus
	unnecessary estimation risk.  We instead desire a pooling of estimated
	variance, toward a single nugget for the entire global surface, although
	return to the potential for greater flexibility in our discussion in Section
	\ref{sec:discuss}.

	One option is maximum likelihood, though the sum of normal log densities
	with mean and variance defined by   \eqref{eq:partvar}.  Unfortunately, we
	have not found this to be computationally tractable.  Instead, we prefer a
	moment-based approach. Recognizing that the minimum possible variance is
	$\tau^2\eta$, we select $\eta$ such that this minimum meets our estimate of
	the variance at a specific point.  For each local expert, we find
	$\widehat{\mathrm{mse}}_k =
	\frac{1}{|X_k|} \sum_{i=1}^{|X_k|} (\hat{y}_i^{(k)} - y_i^{(k)})^2$, where
	$\hat{y}_i^{(k)}$ and $y_i^{(k)}$ are fitted values from model $k$ and
	observed training values, respectively.  We then use the empirical
	$\widehat{\mathrm{mse}} =
	\sum_k
	\widehat{\mathrm{mse}}_k / N_k$ as our estimate of the minimum variance for
	a model, $\hat{s}^2$. This leads to the selection of $\hat{\eta} =
	\hat{s}^2/\hat{\tau}^2$ as the unified nugget for all experts.

	\subsection{Illustration}

	A broader suite of empirical benchmarking results is provided in Section
	\ref{sec:empirical}.  However, before moving to sequential selection of
	experts in Section \ref{sec:seq}, we pause briefly here to provide some
	initial proof-of-concept results in this simpler, static setting. Here, and
	later in Section \ref{sec:empirical}, we contrast model quality and
	performance along several key metrics.  First, to compare quality of
	predictions, we measure the out-of-sample root mean squared prediction error
	(RMSE).  For $N'$ testing data examples,
	\[
	\mathrm{RMSE} = \sqrt{\frac{1}{N'} \sum_{i=1}^{N'} (y_i -
	\mu(x_i))^2}.
	\] 
	Second, we consider the proper scoring rule introduced
	by Eq.~(27) in \cite{gneiting2007scoring}, which is a reduced form of
	the scoring rule utilizing two moments -- a special case of Eq.~(25)
	from that same paper.  
	\[ 
	\mathrm{score}\left(\mu(X), y\right) = -
	\frac{1}{N'} \sum_{i=1}^{N'} \frac{\left(y_i -
	\mu(x_i)\right)^2}{\sigma^2(x_i)} - \mathrm{log}(\sigma^2(x_i)) 
	\]
	Notice how score offers good balance of predictive accuracy through $(y_i -
	\mu(x_i))^2$, and uncertainty quantification through $\sigma^2(x_i)$.  We
	expect that whatever we might lose in predictive accuracy from LAGP can be
	made up by leveraging global information for proper variance estimates.
	Lastly, we note predictive time.  This metric is the main catalyst for
	developing the PALM methodology, and we would expect a massive advantage
	over other methods with a similar predictive accuracy.

	Table \ref{results} summarizes the results of a simulation study comparing
	PALM and LAGP on these important metrics, including time per prediction in
	seconds.  We fit a PALM with $K=100$ space-filling centers, each with the
	default {\tt laGP} settings, to Herbie's Tooth (\ref{eq:herbie}) generated
	on an $N=100 \times 100$ grid in 2d. Training responses are sampled under a
	standard zero-mean Gaussian with $\mathrm{sd}=0.05$.   The testing set is
	generated form the same function on a $101\times 101$ grid, deliberately
	spaced to miss the training data locations.  Observe that RMSE and score are
	practically identical between the two predictors. Even for a small PALM
	(only 100 local models, utilizing fewer than 5000 of the 10000 training
	sites), we have achieved comparable results to LAGP, which utilizes every
	training data location, while predicting in a thirtieth of the time.

	\input{results.txt}

\begin{figure}[ht!]
		\centering
		\includegraphics[width=65mm,trim=0 30 0 10]{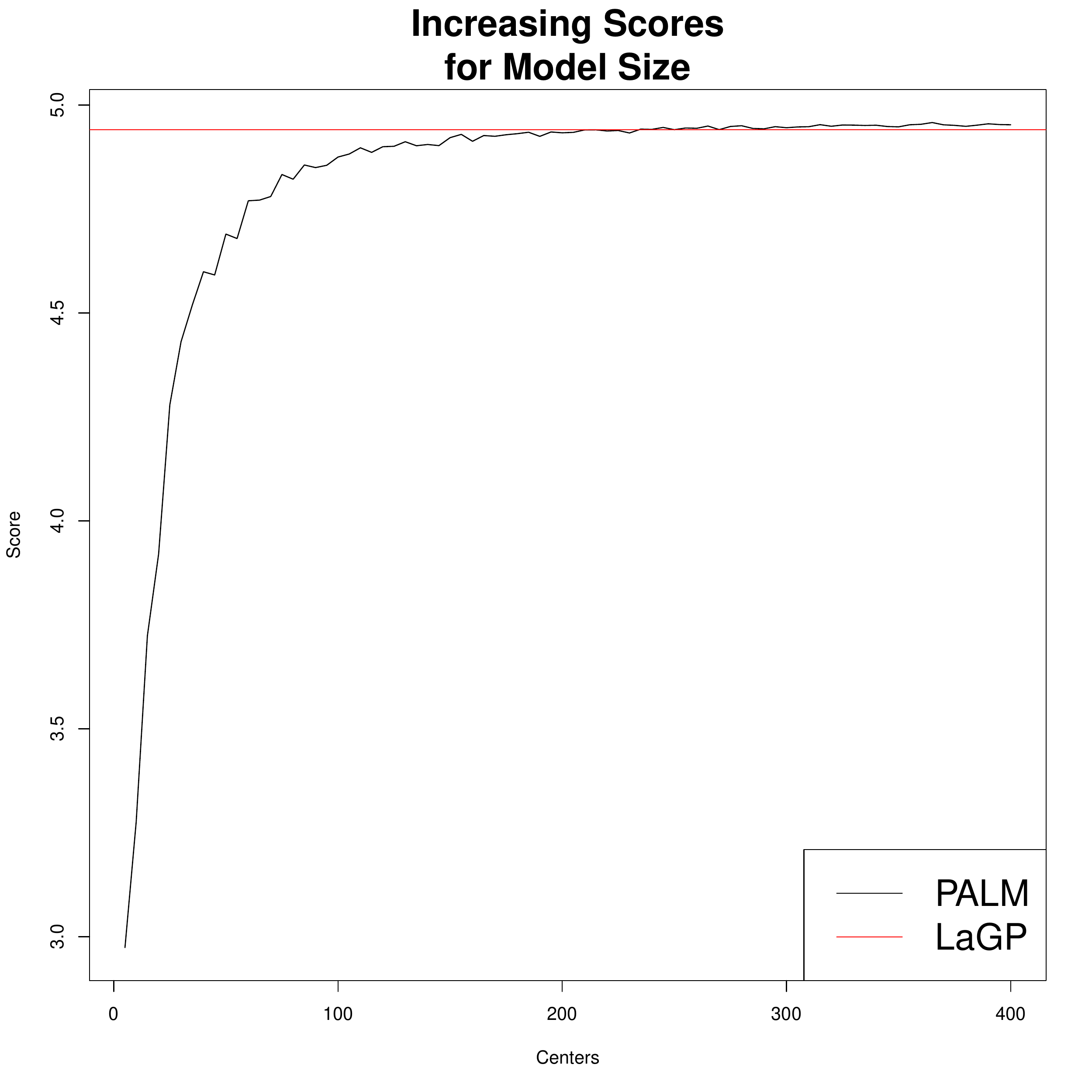}%
		\includegraphics[width=65mm,trim=0 30 0 10]{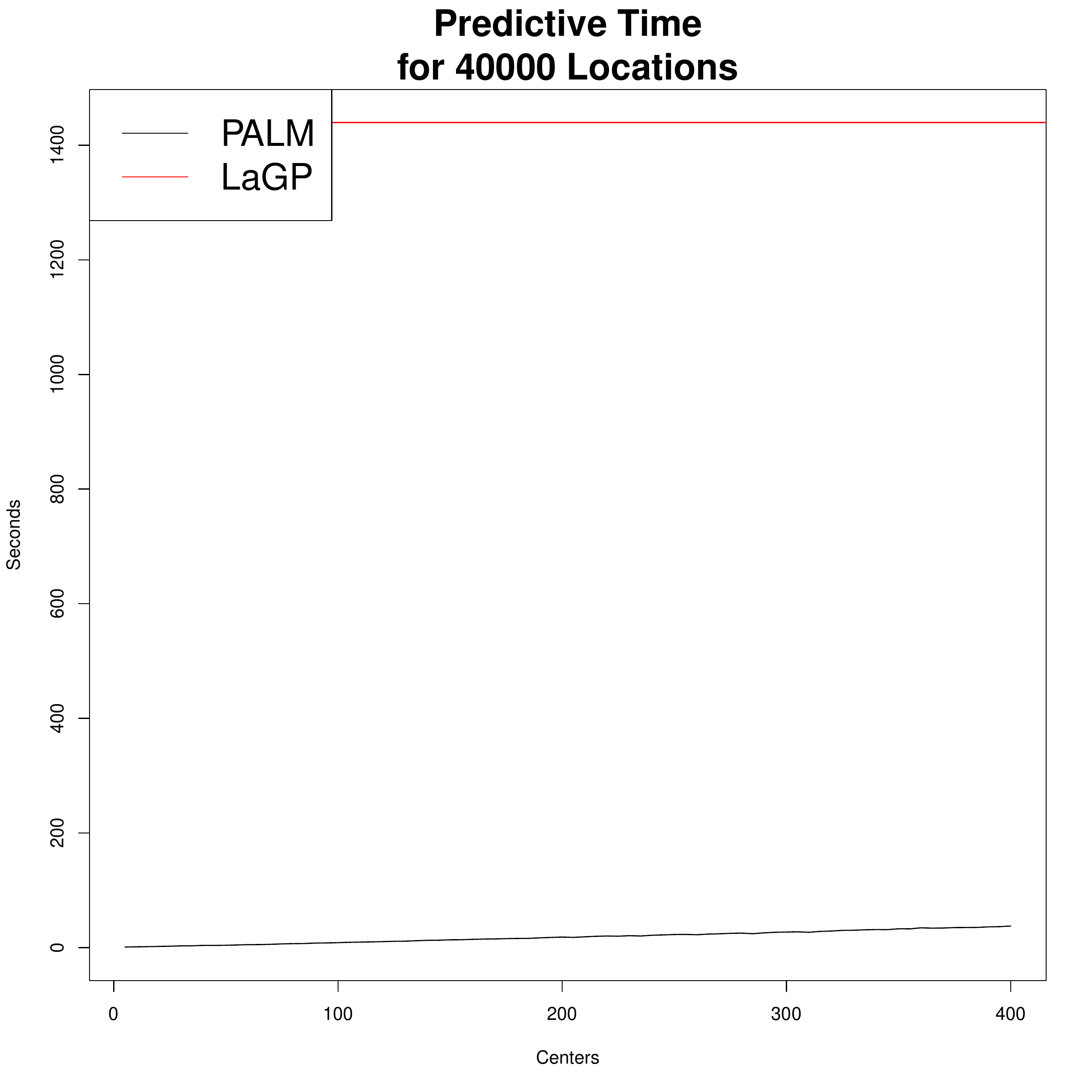}
		\caption{Score (left) and predictive time (right) for increasingly
		large PALM v.~LAGP.}
		\label{f:noise}
	\end{figure}

	If we increase the size of the PALM, we can match the performance of LAGP
	while maintaining predictive advantage. Figure \ref{f:noise} shows the
	predictive capability of PALMs of various sizes versus that of LAGP.  To
	accommodate predictors trained on larger data sets, we increase the size of
	the training data set to 40 thousand on a $N=200\times 200$ grid, and
	similarly expanded the testing set is on a $199\times 199$ grid (size
	39601), both otherwise generated as described above.  The black line in the
	right panel of the figure represents the score of PALMs with the number of
	space-filling centers indicated on the $x$-axis. On the right, total
	predictive time is plotted for the same fits.  LAGP is in
	red  on both plots.  Note that the line is horizontal because LAGP
	builds a separate model for each prediction, and thus does not have an
	increasing size component. Of particular interest is PALM's predictive time,
	shown in the right panel, which indicates a very slow increase.  Despite
	diminishing returns in score as model complexity increases, a larger PALM
	provides generally better performance in terms of cost--benefit trade-off
	relative to raw LAGP.

\begin{figure}[ht!]
		\centering
		\includegraphics[width=65mm,trim=0 30 0 10]{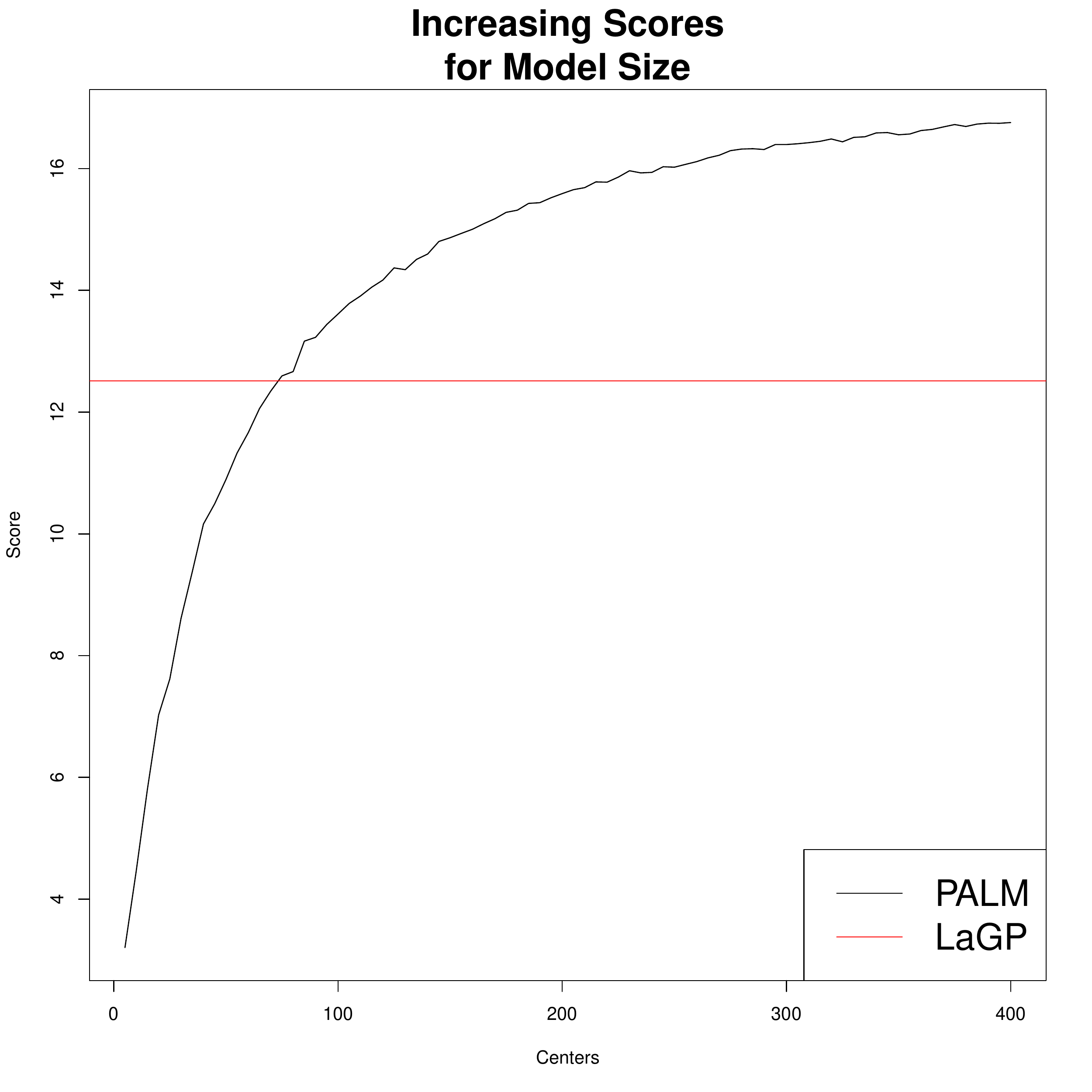}%
		\includegraphics[width=65mm,trim=0 30 0 10]{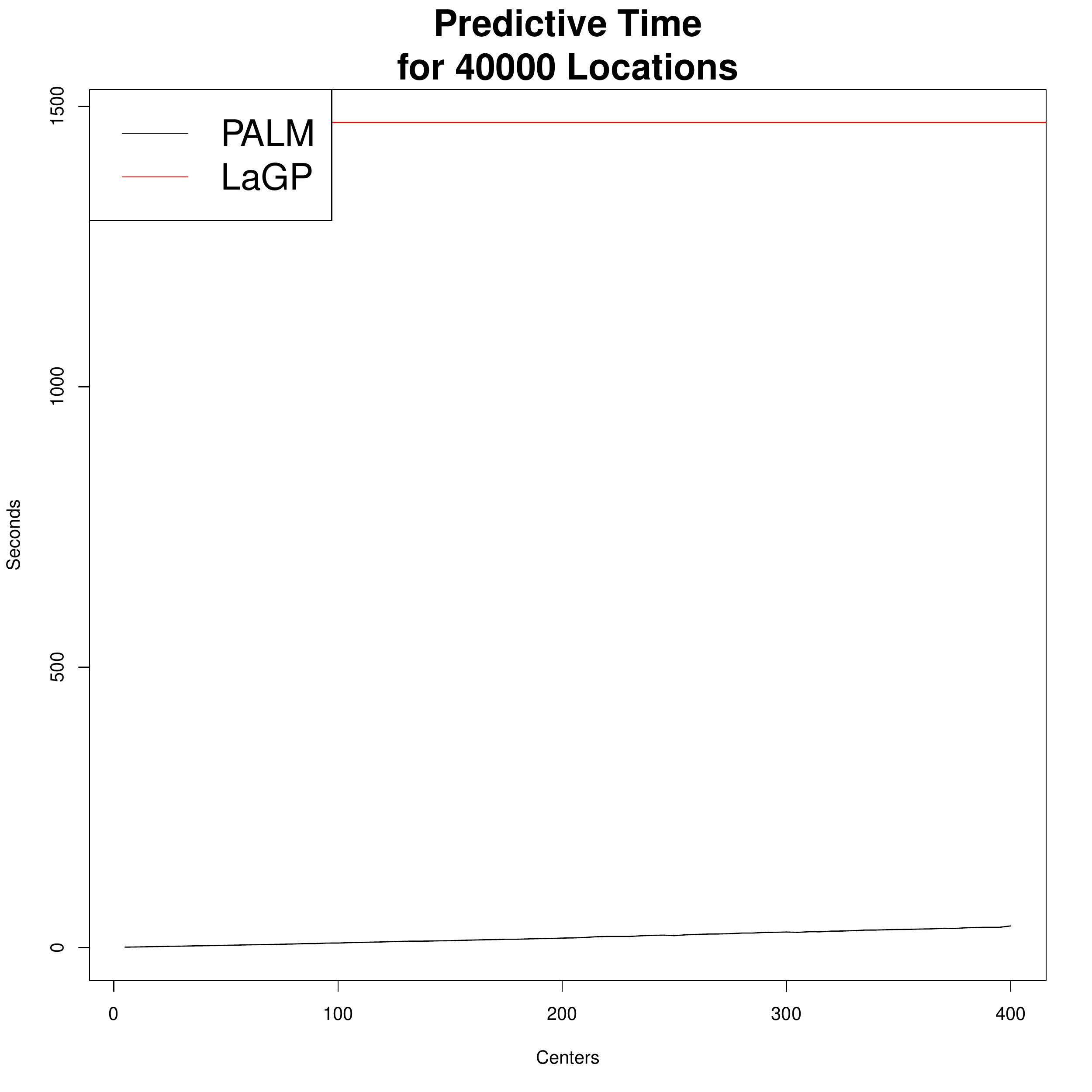}
		\caption{Analog of Figure \ref{f:noise} on
		deterministic Herbie's tooth.}
		\label{f:2d}
	\end{figure}

	For deterministic data, accurate variance representation allows for
	PALM to perform much better than LAGP with a vastly improved predictive
	time.  Figure \ref{f:2d} shows the performance of both models on a
	Herbie's tooth \eqref{eq:herbie}, generated here with a standard
	deviation of 0.  The training and test sizes, and design locations
	remain the same as above. The plot should be read the same way as
	Figure \ref{f:noise}.  Notice that PALM matches and surpasses the
	performance of laGP while maintaining a large advantage in time.

	\section{Sequential Selection} \label{sec:seq}

	Until this point we have utilized equally-spaced experts (or LAGP centers)
	when PALM fitting. While this suffices for low-dimensional examples
	exhibiting highly regular (i.e., stationary) dynamics, bigger problems and
	more complex (e.g., nonstationary) response surfaces may benefit an uneven
	spread of computational resources to accurately capture the global
	surface.  Consider the 2d exponential function from
	\cite{gramacy2008bayesian}.
	\begin{align}
	f(x) = x_1 \exp\{-x_1^2 - x_2^2\} 
	\label{eq:glee} 
	\end{align} 
	Exponential decay means that the response surface is essentially zero
	everywhere except near the origin.  Although dynamics are smooth, bumps near
	the origin create subtle nonstationarity.  Capturing both bumpy and flat
	regions, we take inputs in $[-2,6]^2$.

\begin{figure}[ht!]
		\centering
		\includegraphics[width=.5\textwidth,trim=0 40 0 0]{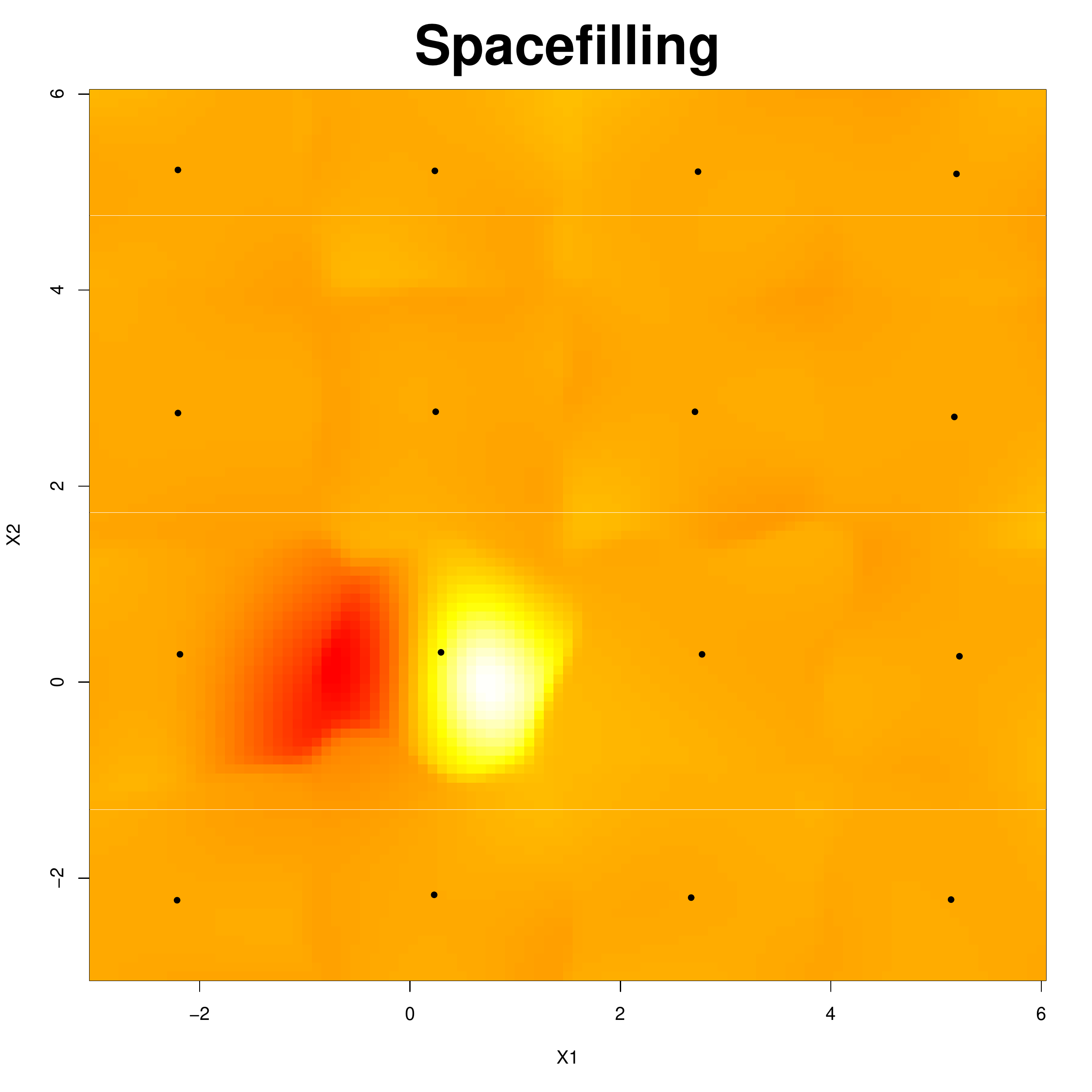}%
		\includegraphics[width=.5\textwidth,trim=0 40 0 0]{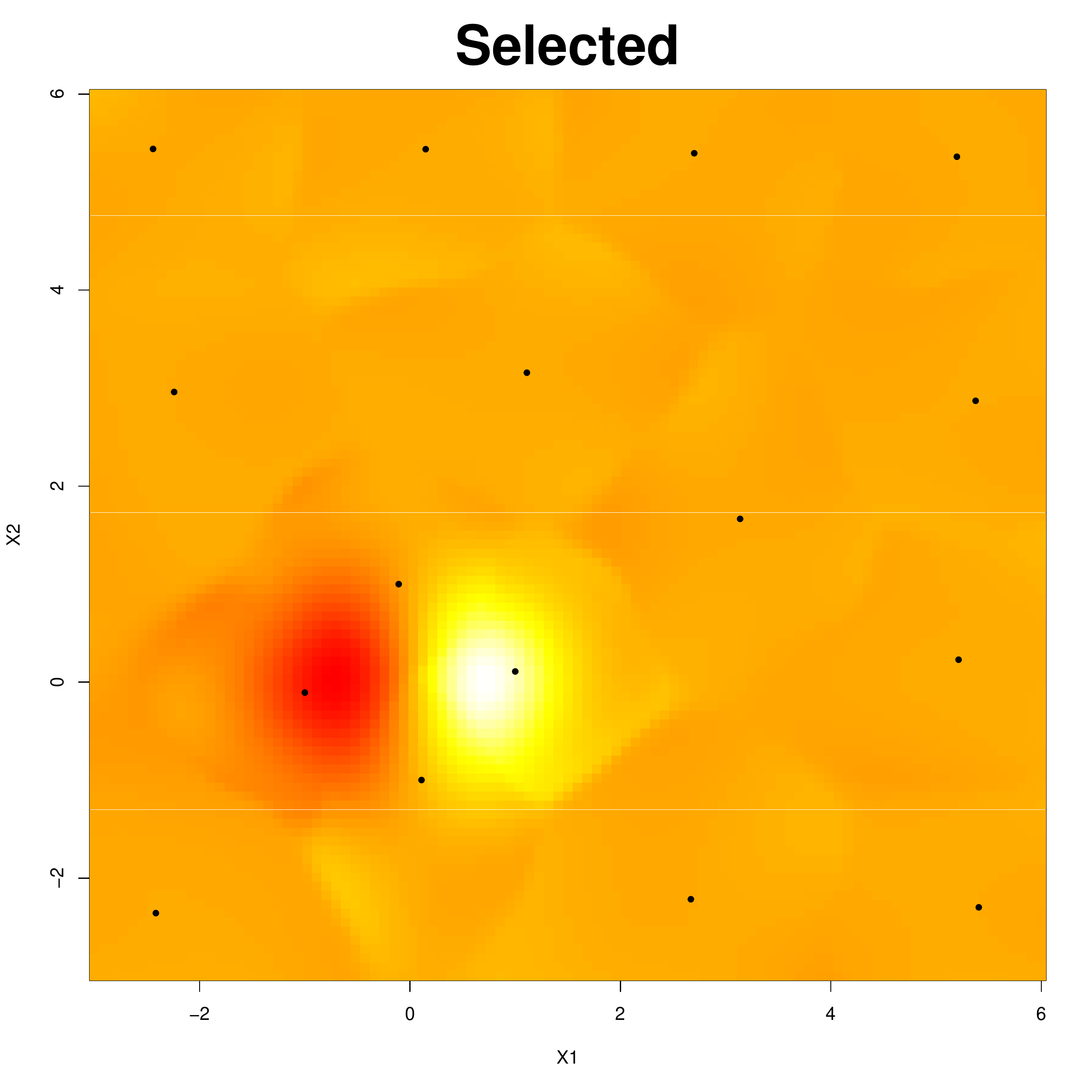}
		\caption{Two different schemes for location of local experts: gridded (left) and re-focused on the interesting region (right).}
		\label{selection}
	\end{figure}

	In Figure \ref{selection}, we have intentionally limited the number of local
	experts to illustrate how their uneven distribution may improve global
	accuracy.  In the left panel, 16 experts have been assigned to fill out the
	space.  Observe that only one local expert center resides in the interesting
	region near the origin.  In the right panel, four points have been assigned to
	represent the interesting area before the other 12 are filled in by a space
	filling scheme in the flat area of the function.  This adjustment in scheme
	results in much better coverage of the more complicated area, and an overall
	reduction in error, as we shall quantify in more detail in due course.  For
	now, notice that the right panel exhibits greater relative isotropy compared
	to the left panel.  Looking at Eq.~(\ref{eq:glee}), radial decay is evident
	in $x_1 \times x_2$ space.

	\subsection{Greedy selection of PALM centers}
	Allocation of experts/centers in the simple case  above is rather
	straightforward if one has prior knowledge of the surface, which is in
	general unrealistic. For most functions we will not have prior knowledge of
	hard to compute areas or, in more than 2 dimensions, an ability to
	directly visualize them.  For similar reasons, optimizing the locations
	of all centers simultaneously could be challenging.  Instead, we find it
	advantageous to proceed sequentially. Given an existing PALM, we consider
	optimization of the selection of one additional local expert.  Such a greedy
	strategy, targeting areas of the input space demonstrating the greatest
	benefit to expanded modeling fidelity, mirrors LAGP's scheme for choosing
	local design sites sequentially, e.g., using a variance reduction scheme
	(i.e., ALC).

	In our context of greedy center selection we have the luxury of measuring
	out-of-sample accuracy directly, rather than relying on a model-based
	deduction of predictive uncertainty (ALC).  Namely, we can calculate
	residuals between predicted and actual responses under the PALM fit.
	Therefore, our development of criteria for greedily selecting new centers
	focuses on identifying areas of the input space suffering from large such
	(absolute) residuals.  Specifically we deploy $k$-means
	\cite{lloyd1982kmeans} on pairs $(x,r)$, where $r = |y-\hat{y}|$ and $x$
	is the input location(s) of those $y$-values, in order to find clusters of
	high residuals where additional centers may be most helpful at increasing
	accuracy.  In practice we find it helpful to scale those absolute
	residuals to be commensurate in size with the coding of inputs $x$ in
	order to encourage the algorithm to form spatially contiguous clusters.  A
	PALM with $K$ centers should have on the order of $K$ ``local maxima'' for
	poorly predicted areas, so in practice we set the number of $k$-means
	clusters to be equal to the number of local experts in the model.  We
	denote the $k$-means algorithm, applied to a data set, $X_N$, with $k$
	clusters as kmeans$(X_N,k)$. Next, we identify the cluster with highest
	average residual, place a bounding box around that geographical region of
	the input space, and perform a local numerical optimization in order to
	fine-tune the location of the new center.

	\begin{figure}[ht!]
		\centering
		\includegraphics[width=.6\textwidth]{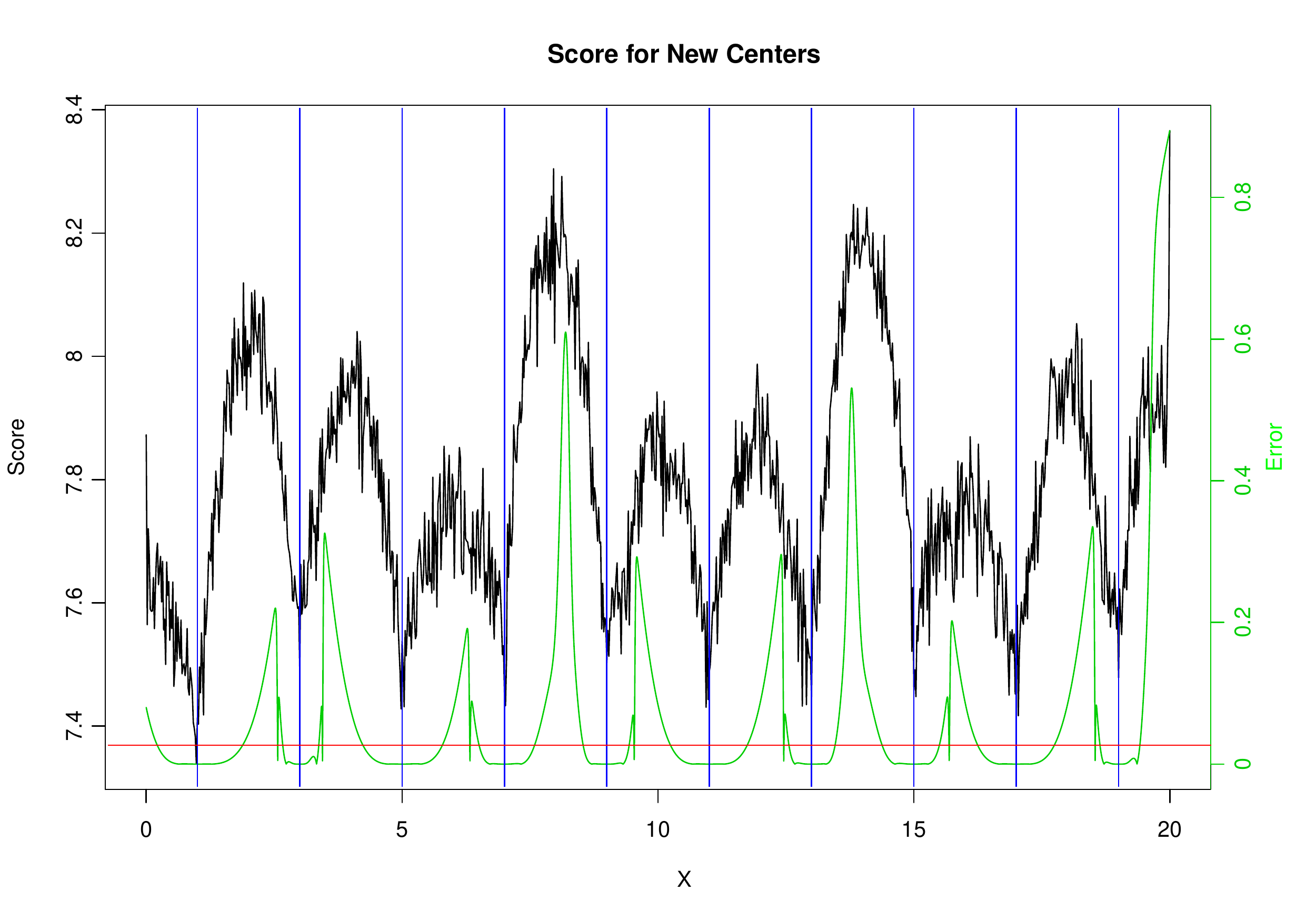}
		\caption{Score for potential new local expert placement against
		the absolute error of the existing model. The horizontal red line
		represents the score of an existing model, while the black line represents
		the score for a model updated with a center located at that spot along the
		number line.  The (right) ``error'' axis indicates the absolute error for
		predictions using the existing model at that location.}
		\label{fig:modes}
	\end{figure}

	Our choice of that final criteria for local optimization is nuanced. Ideally
	we would maximize score or aggregated absolute residuals.  However, as a
	function of LAGP sub-design(s), both are discontinuous -- score
	pathologically so -- thwarting off-loading of optimization to simple
	library-based schemes. Instead, we leverage an empirical observation that
	good predictive areas tend to be far away (in the input space) from other
	local models in hard-to-predict areas.  In other words, space-filling in a
	high residual region -- e.g., as selected by $k$-means -- offers a decent
	surrogate for score or residual-optimizing. Figure \ref{fig:modes} shows the
	intuition for this idea.  The horizontal red line represents the score of a
	PALM designed to fit a grid of 10,000 equally-spaced points along a sine wave
	with centers located at the vertical blue lines.  The jagged black line is
	the score for a new center placed at that spot on the $x$-axis.  The green
	line is the absolute residual at that point in the existing PALM.

	\begin{algorithm}[ht!]
		\begin{algorithmic}[1]
			\REQUIRE existing {\tt PALM} model with $N$ local experts, input locations
			$X$, response vector $y$, matrix of existing center
			locations $C$
			\STATE $\hat{y} \leftarrow$ predict$({\tt PALM}, X)$
			\hfill
			\COMMENT{Predict on the known input locations}
			\STATE r $\leftarrow$ $|y-\hat{y}|$
			\hfill
			\COMMENT{compute absolute residuals}
			\STATE $R \leftarrow$ bind$(X,r)$
			\hfill
			\COMMENT{Combine input space and residuals}
			\STATE $K \leftarrow \mathrm{kmeans}(R, N)$
			\hfill
			\COMMENT{Form into $k$ contiguous clusters}
			\STATE $X_k \leftarrow K[\bar{r} = \max(\bar{r}),-r]$
			\hfill
			\COMMENT{Find the cluster with high residuals}
			\STATE $B \leftarrow$ apply$(X_k,$ range$)$
			\hfill
			\COMMENT{Compute the bounding box in the input space}
			\FOR{$i$ in $1$ \TO $M_s$}
			\STATE $c_{\mathrm{start}} \backsim$
			uniform$\left(B\right)$
			\hfill
			\COMMENT{Uniformly generate starting point}
			\STATE $s_i \leftarrow$ optim$_{c|c_{\mathrm{start}}
			\in B} \min \lVert c-z \rVert_{z \in C}$
			\hfill
			\COMMENT{Optimize from the random start}
			\ENDFOR
			\STATE $c_{\mathrm{new}} \leftarrow \mathrm{argmax}_{s
			\in S} \min \lVert s-z \rVert_{z \in C}$
			\hfill
			\COMMENT{Find overall best solution}
			\STATE ${\tt PALM} \leftarrow {\tt PALM} \cup {\tt
			laGP}(c_{new})$
			\hfill
			\COMMENT{Append the {\tt PALM} with a new {\tt laGP}}
			\STATE $C \leftarrow C \cup c_{\mathrm{new}}$
			\hfill
			\COMMENT{Add in the new center location}

		\end{algorithmic}
		\caption{Computing the location of one more center given an existing PALM
		fit.}
		\label{alg:seq}
	\end{algorithm}

	Observe in the figure that if we were to select a local area that had high
	absolute error, and then subsequently choose a space filling
	location/midpoint within that space, it would correspond with a high
	score/high absolute residual area of the input space.  Also note that the
	only center locations which make the PALM worse than the existing model are
	the locations of the current centers. Therefore we have designed our greedy
	scheme to choose a new center, $c$, such that $c = \mathrm{argmax}_{c \in B}
	\min \lVert c-z \rVert_{z \in C} $ where $c$ is the location of the new
	center to be added, $B$ is the bounding box around the poorly predicted
	cluster of residuals, and $C$ is the matrix containing the locations of
	centers already in the model appended by the corners of the bounding box.  This ``maximin'' criterion is a continuous
	function which is easily optimized with library methods such as {\tt optim}
	in {\sf R}, although within the bounding box there may be several local
	maxima. To combat this we engage a randomized multi-start scheme. Finally,
	build an LAGP forming local expert for the PALM centered at the solution.
	For concreteness, the complete sequential PALM center-updating algorithm,
	with $M_s$ being the number of multi-start maximin attempts is summarized 
	in Algorithm \ref{alg:seq},

	\subsection{Illustration}

\begin{figure}[ht!] 
		\centering
		\includegraphics[width=0.38\textwidth]{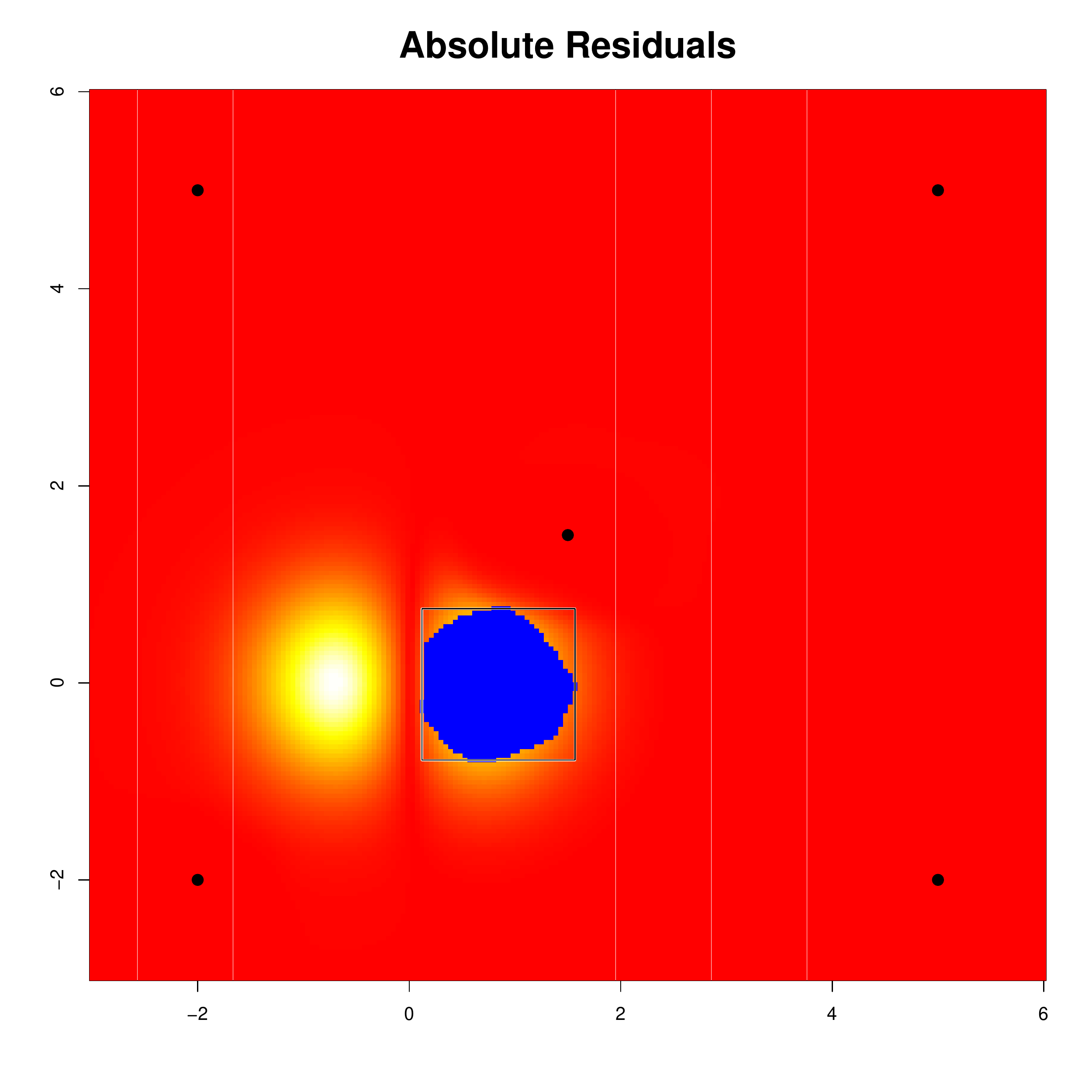}%
		\includegraphics[width=0.38\textwidth]{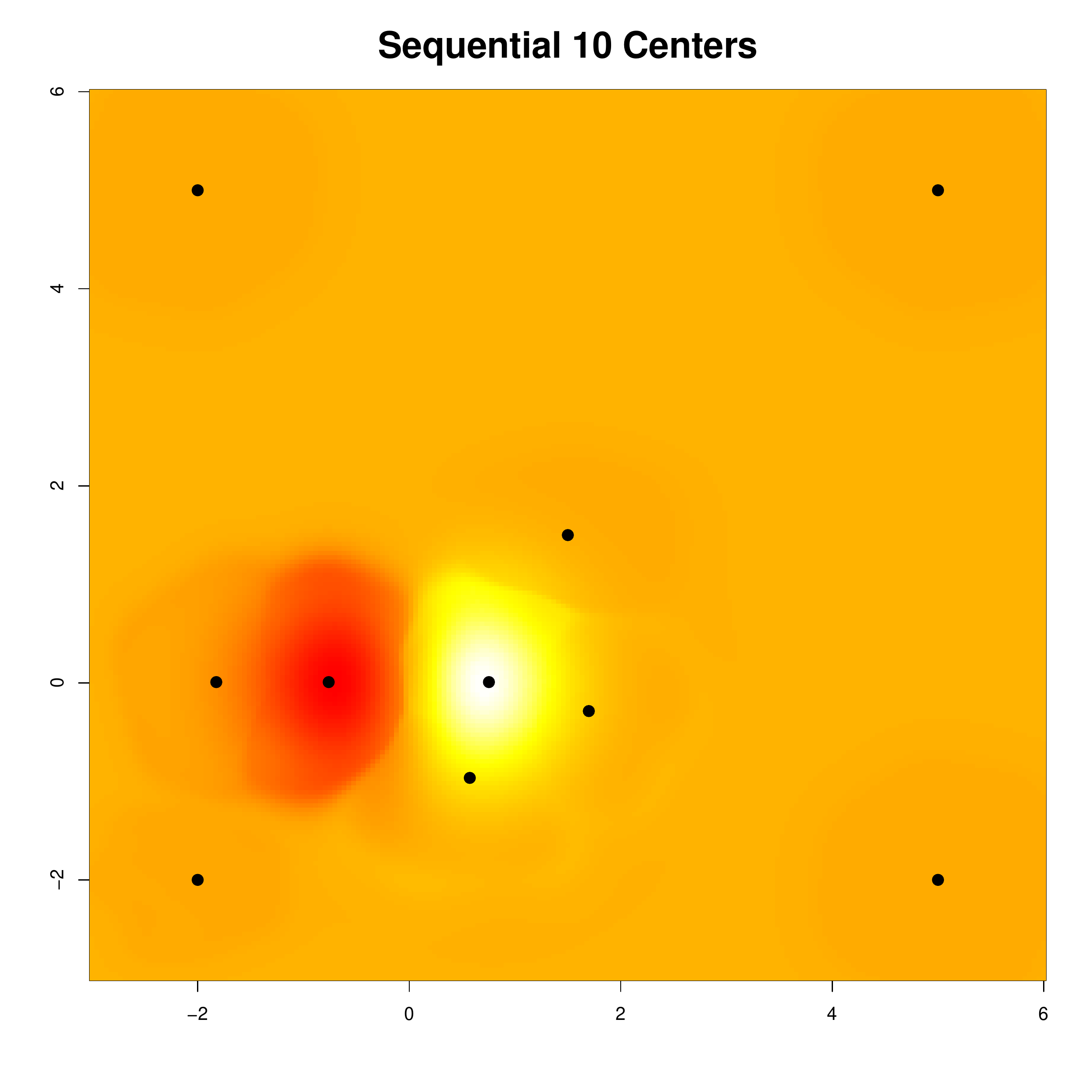}\\
		\includegraphics[width=0.38\textwidth]{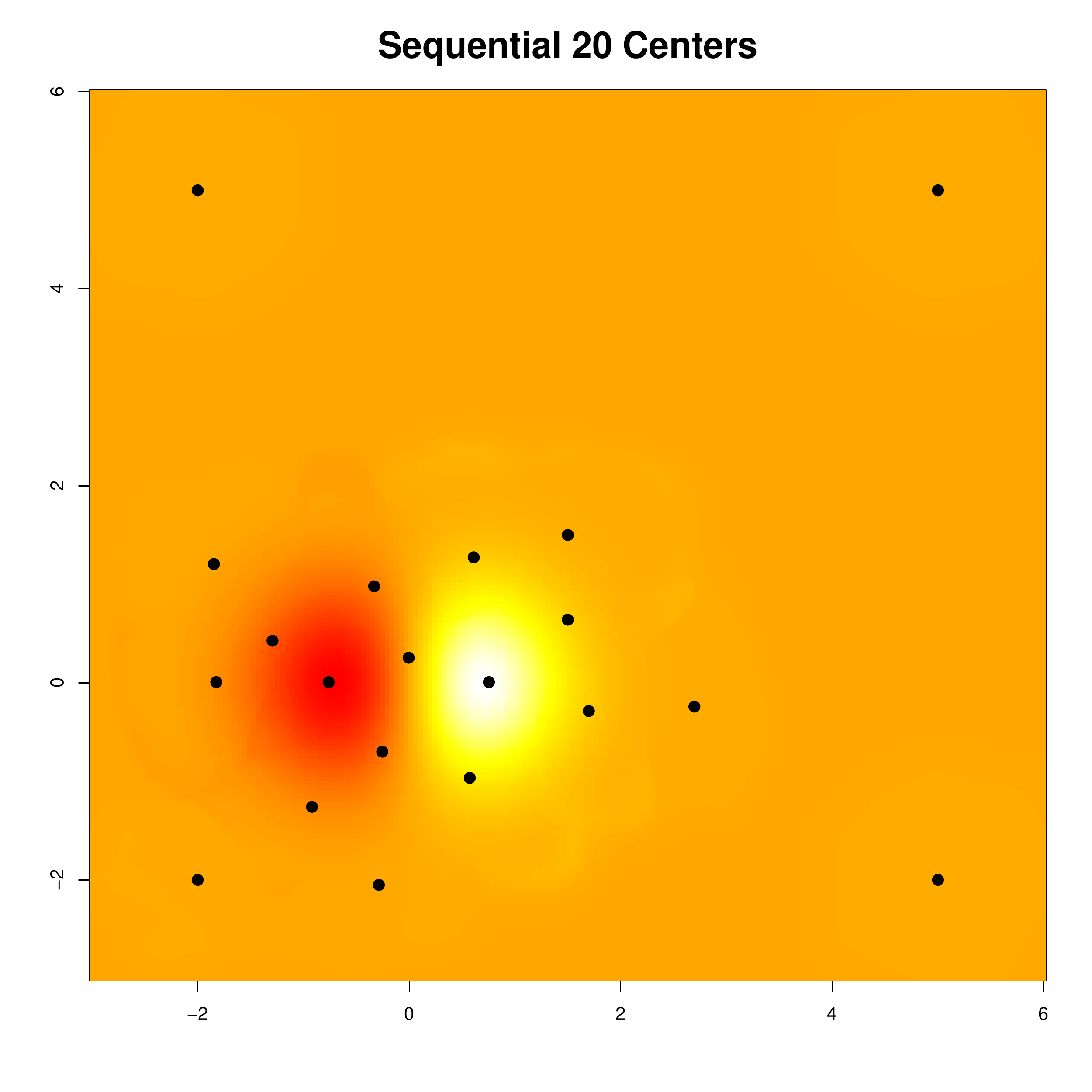}%
		\includegraphics[width=0.38\textwidth]{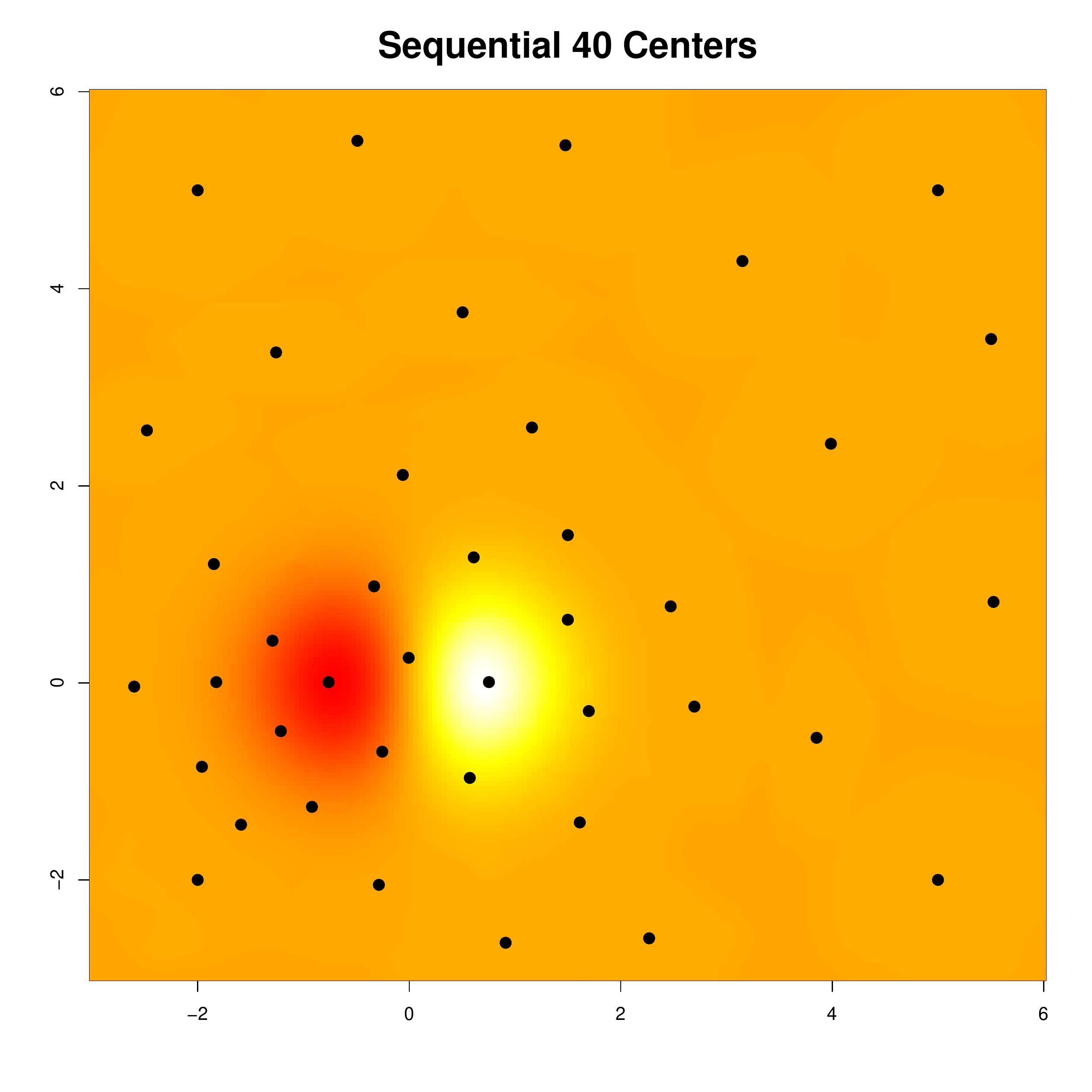}
		\caption{An illustration of the sequential selection process.
		The top left panel has 5 space filling centers.  The top right,
		bottom left, and bottom right follow our sequential algorithm
		to achieve total sizes of 10, 20, and 40 centers respectively}
		\label{fig:seq_select}
	\end{figure}

	To get familiar with this greedy scheme, we provide a visualization and
	empirical comparison on the Gramacy and Lee function
	\eqref{eq:glee}.  A regular grid of $N=40000$ training data pairs is
	created deterministically.  We begin with a PALM composed of a small space
	filling design with $K=5$ centers.  The absolute residuals from this model
	are seen in the top left of Figure \ref{fig:seq_select}.
	None of the points have been placed within the area of interest, resulting
	in two high areas surrounded by relative flatness.  The k--means algorithm
	selects one area of high residuals, shown here in blue, and a bounding box
	is drawn around them.  In this case, where there are no existing centers
	inside the box, the new expert will be designed to predict exactly the
	middle of the high residual area. To the right, we have placed five more
	centers, one at a time following the greedy scheme outlined in Algorithm
	\ref{alg:seq}.  Observe that all have been chosen in the difficult area of
	the input space, near the origin.  The bottom left panel shows a further
	10 sequentially selected centers.  Now our greedy selector is starting to
	form a good picture of the interesting region and is exploring putting
	centers around the edges to determine where the function returns to a flat
	state. By the time we reach $K=40$ centers, in the bottom right, the PALM
	has well described the whole region and is selecting additional centers in
	a space filling fashion around the interesting region as well as
	elsewhere.

	\begin{figure}[ht!] \centering
		\includegraphics[width=.5\textwidth]{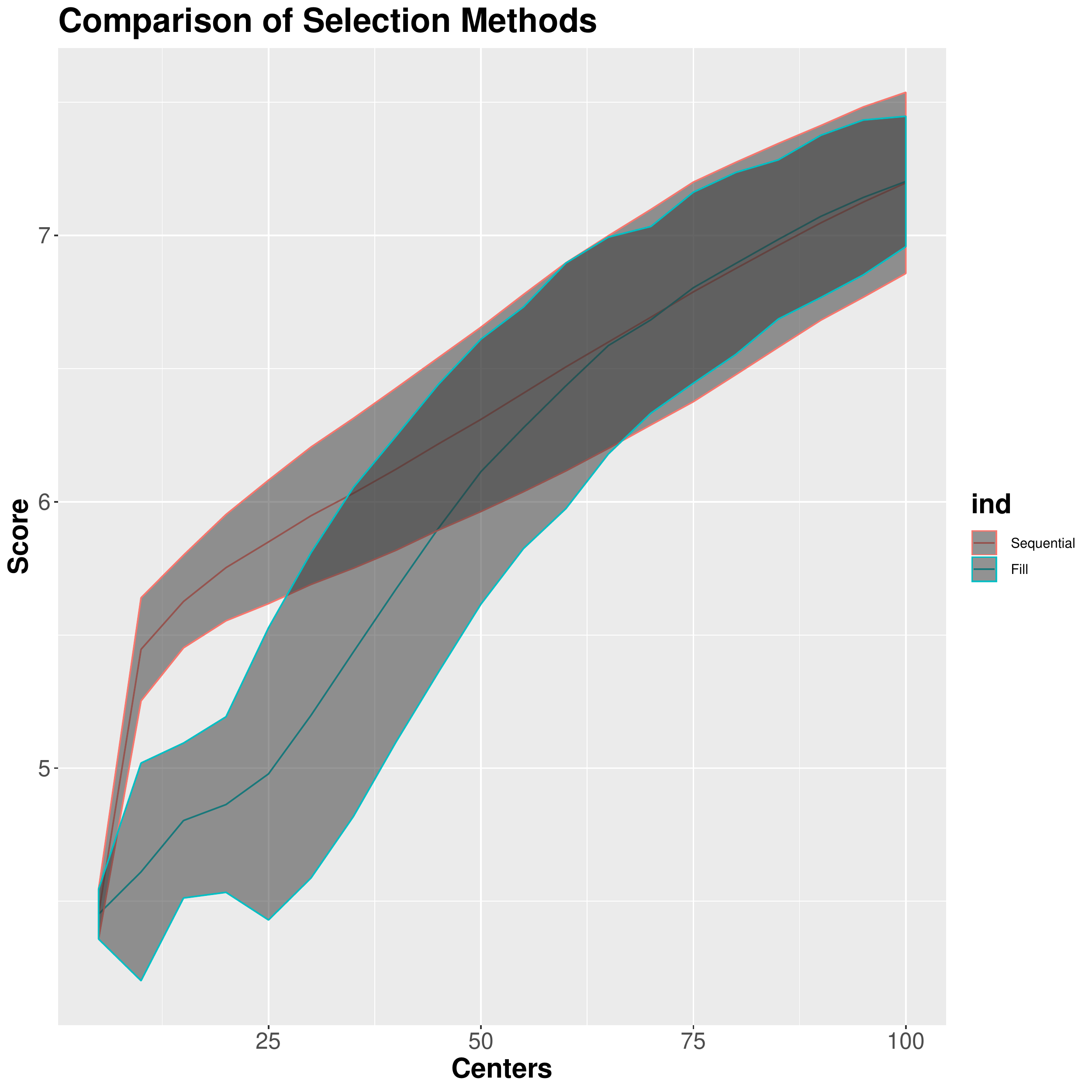}
		\caption{The score of models of various sizes when space
		filling, or applying a sequential selection algorithm on the Gramacy and
		Lee function.}
		\label{fig:sequence}
	\end{figure}

	Figure \ref{fig:sequence} shows how sequentially selecting can potentially
	drastically improve PALM fits, but will do no worse than a space filling
	baseline.  This time we generated commensurately-sized data
	\eqref{eq:glee} with noise (sd $=0.01$).  For 100 such randomly generated
	data sets, PALMs were built with 5 to 100 centers incrementing by 5 in a
	space-filling fashion.  The out-of-sample performance of this
	space-filling comparator is indicated by the blue lines in the figure. The
	red line, however, represents the above greedy/sequential selection
	scheme.  An initial/seed spacefilling model of size 5 was built, and all
	subsequent centers were added by greedy selection following Algorithm
	\ref{alg:seq}.  Notice that the sequential scheme has a sharp spike for
	smaller models, but levels out to match the space filling models in the
	long run.  This represents a early concentration of centers around the
	``interesting area'' near the origin before the sequential scheme starts
	filling in the rest of the space.

	In general, if there is no prior knowledge of the interesting areas of
	a function, sequential selection of center locations will provide a
	potential boost in model accuracy by concentrating resources in the
	active subspace of the model.  When the resources allocated are enough
	to describe the entire surface well, a sequentially selected model will
	match the performance of a spacefilling scheme.  By front loading
	calculation on determining center locations, the overall performance of
	a PALM fit can be improved.


	\section{Empirical work} \label{sec:empirical}
	
	To further test the viability of the PALM framework, passages below describe
	its application on more complicated data sets than the illustrative examples
	given thus far.  First we shall demonstrate the effectiveness of our method
	on a real data example from a spatial data forecasting competition, allowing
	comparisons to be drawn to a variety of other GP-based and related
	large-scale smoothing methods.  Then we show a comparison of space filling
	versus sequentially selected center designs on a classic three dimensional
	test function.
	
	\subsection{Satellite Data}

\begin{figure}[ht!] \centering
		\includegraphics[width=.5\textwidth]{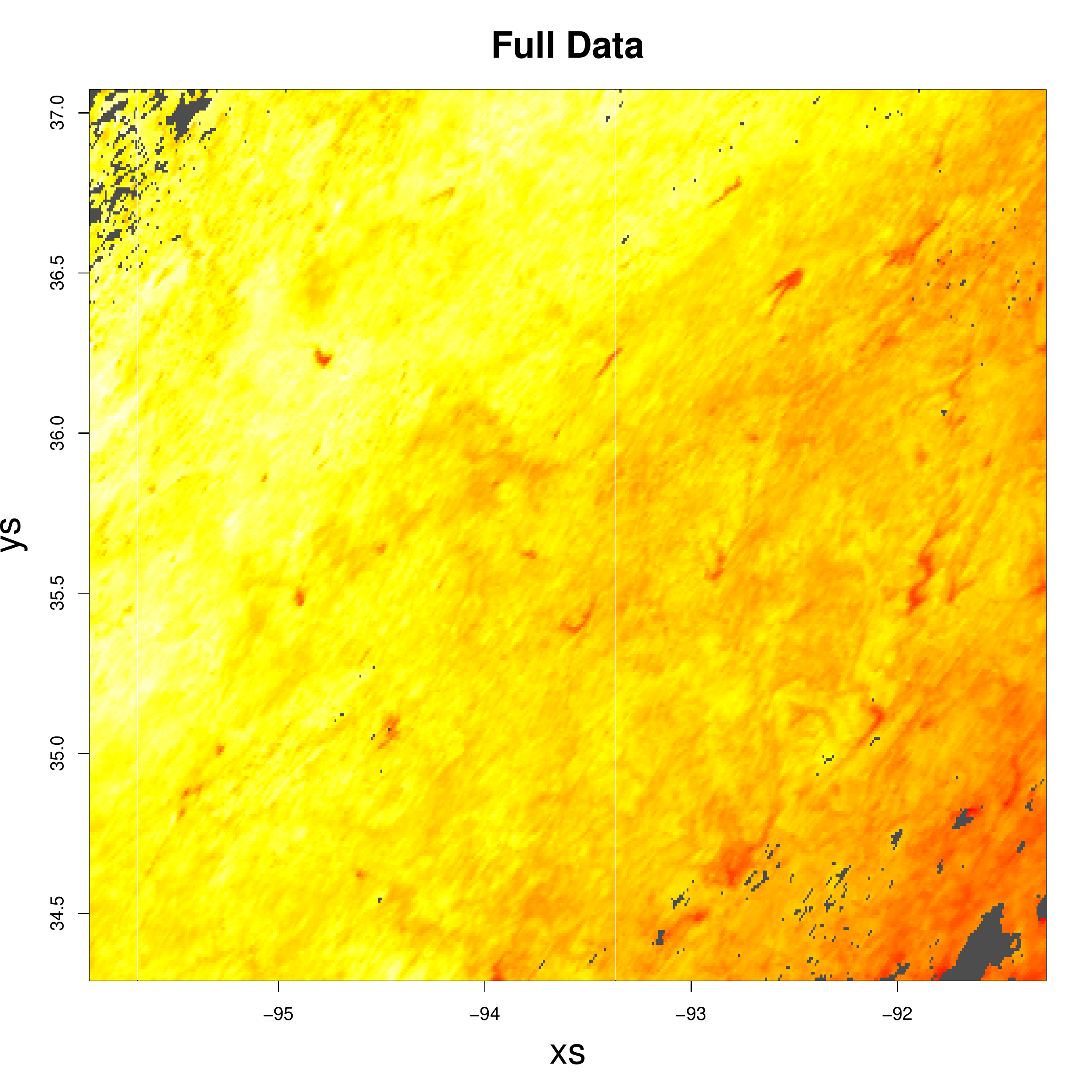}%
		\includegraphics[width=.5\textwidth]{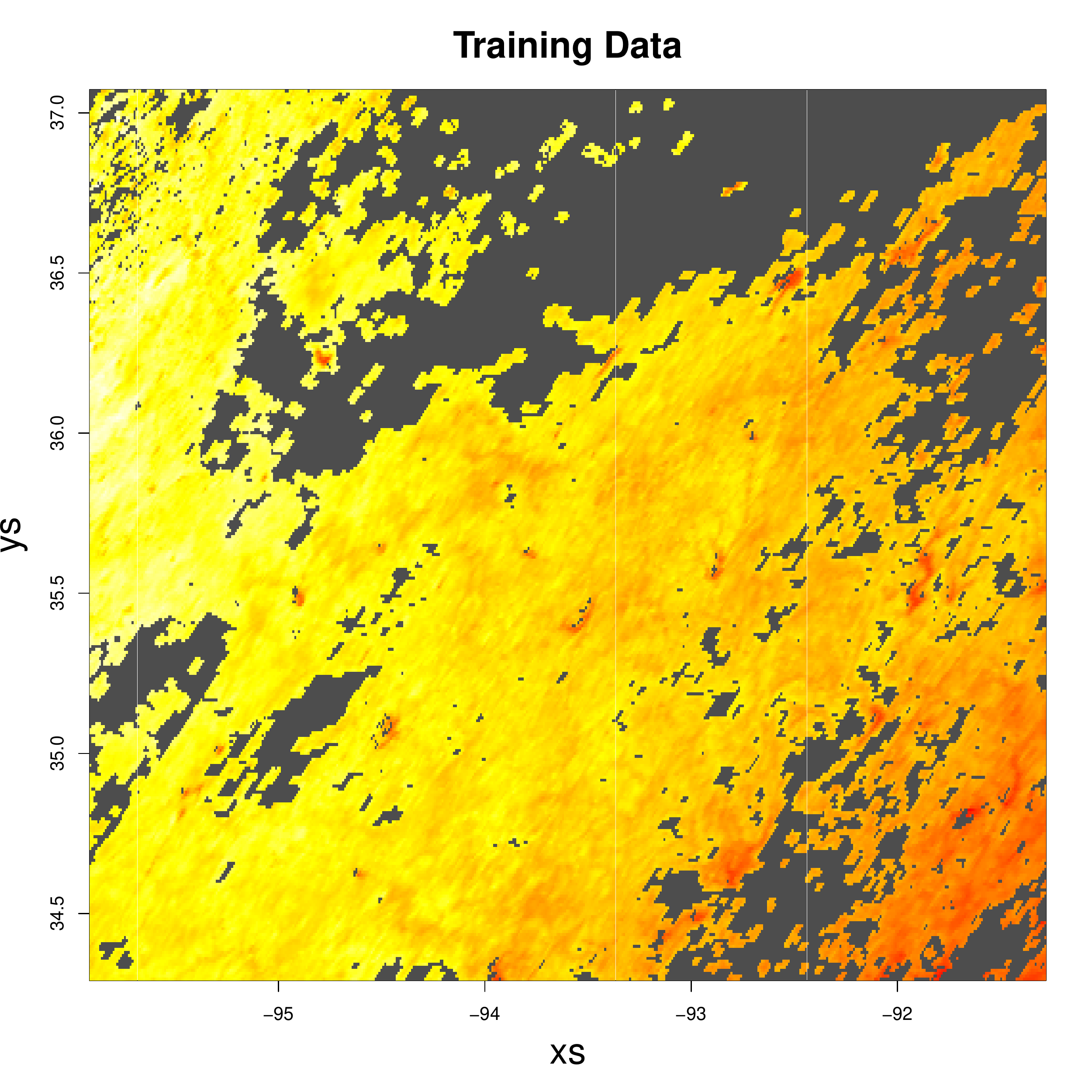}
		\vspace{-0.4cm}
		\caption{Left: full satellite temperature data set.  The few missing
		values are displayed in gray.  Right: dataset subset used for training.
		Much more data is left intentionally missing.}
		\label{fig:temps}
	\end{figure}

	In \cite{heaton2019case}, thirteen state-of-the-art spatial smoothing
	methods, many based on GPs, were compared out-of-sample on satellite
	temperature data under a variety of performance metrics. This provides a
	real data example over which we can benchmark PALM to modern
	competitors.   Several were reviewed in Sections
	\ref{sec:intro}--\ref{sec:partagg}, including a variation on LAGP.  All
	methods and their implementation can be found in the repository linked
	below.
	\begin{center}
	\url{https://github.com/finnlindgren/heatoncomparison}
	\end{center}
	The dataset consists of land surface temperatures measured remotely over
	150 thousand locations selected from a grid, however 1691 record
	``missing'' (or {\tt NA}) values, leaving 148,309.  See the left panel of
	Figure \ref{fig:temps}. \citeauthor{heaton2019case} partitioned these data
	into training and testing sets; see the right panel.  The training data
	represents a large portion of the full data, at 105,569 sites, but has
	substantial swaths of sparse or entirely missing coverage.  In the
	exercise reported in that paper, the testing set was completely hidden
	from competitors.
	
	We built two variations on PALM to add into the mix.  The first uses $K =
	2000$ LAGP experts built in the usual fashion: to directly predict the
	response.  Centers for those models are selected as a space-filling subset
	of the desired prediction locations -- the testing set -- which focuses
	the majority of our computational power on sites where the predictor will
	be scored.  In this way, the application of PALM here is similar to the
	transductive spirit of LAGP \citep{vapnik2013nature}, where learning is
	focused on locations where predictions will be tested.  It should be noted
	that the resulting PALM is not appropriate to describe the entire surface,
	but a prediction-focused model is better suited to the uniquely
	extrapolative nature task at hand. The second PALM variation is fit in two
	stages: first a global GP model is trained to a 1,000-sized sub-sample of
	the data points; then a PALM is fit to residuals from that global subset
	GP using the same 2,000 central locations from the first PALM variation.
	
	Table \ref{t:sat} shows a comparison to the methods from the
	\citeauthor{heaton2019case}~bakeoff on the same metrics given in the
	original paper: mean absolute error (MAE), root mean squared error (RMSE),
	continuous ranked probability score (CRPS; see Section 4.2 of
	\cite{gneiting2007scoring}), interval score (INT; see Section 6.2 of
	\cite{gneiting2007scoring}), prediction interval coverage (CVG), and run
	time (minutes).  Additionally we added a proper scoring rule \citep[i.e.,
	Eq.~(27) from][]{gneiting2007scoring} metric to enable a more direct
	comparison to LAGP connecting to other criteria and comparison made
	throughout this article. Notice that our PALM was run on just one core,
	whereas several of the other methods leverage symmetric multi-core
	parallelization. As discussed previously, and again in Section
	\ref{sec:discuss},  PALM is easily parallelizable. Speculatively, our
	method could be 40 times faster if given access to the same number of
	cores as, say, LAGP.

\input{temps.txt}

	Observe in the table that PALM is performing favorably compared to other
	state-of-the-art methods.  Using only one core, we are able to make
	predictions faster than most other models, and ones which are competitive
	on accuracy (i.e., via MAE and RMSE) of the best of the others.  
	In this example, the ``Global+PALM'' method does slightly better on all predictive measurements. This could be due to the extrapolatory nature of the problem at hand.  With the majority of the computational power focused on an area with relatively little observed data, some estimate of a global trend improves the overall prediction quality.	
	
	While PALM has a speed advantage over many competitors in the
	group, it is in the middle of the pack for most predictive metrics.
	Perhaps this could be improved by increasing the number of local experts,
	$K$ at the expense of time.  Also note that PALM, in its conception, is
	designed to leverage the wide area of accuracy of individual LAGP models
	across the input data (i.e. not to cover large areas of extrapolation).
	Despite this the ``off the shelf'' version of PALM is still competitive
	with state of the art methods.  We have not explored much the effects of
	inverse variance weighting in extrapolation, or the performance of
	``local'' experts with data that is far removed from the central
	prediction location.  
	
	\subsection{Michalewicz Function}
	
\begin{figure}[ht!] \centering
		\includegraphics[width=.5\textwidth]{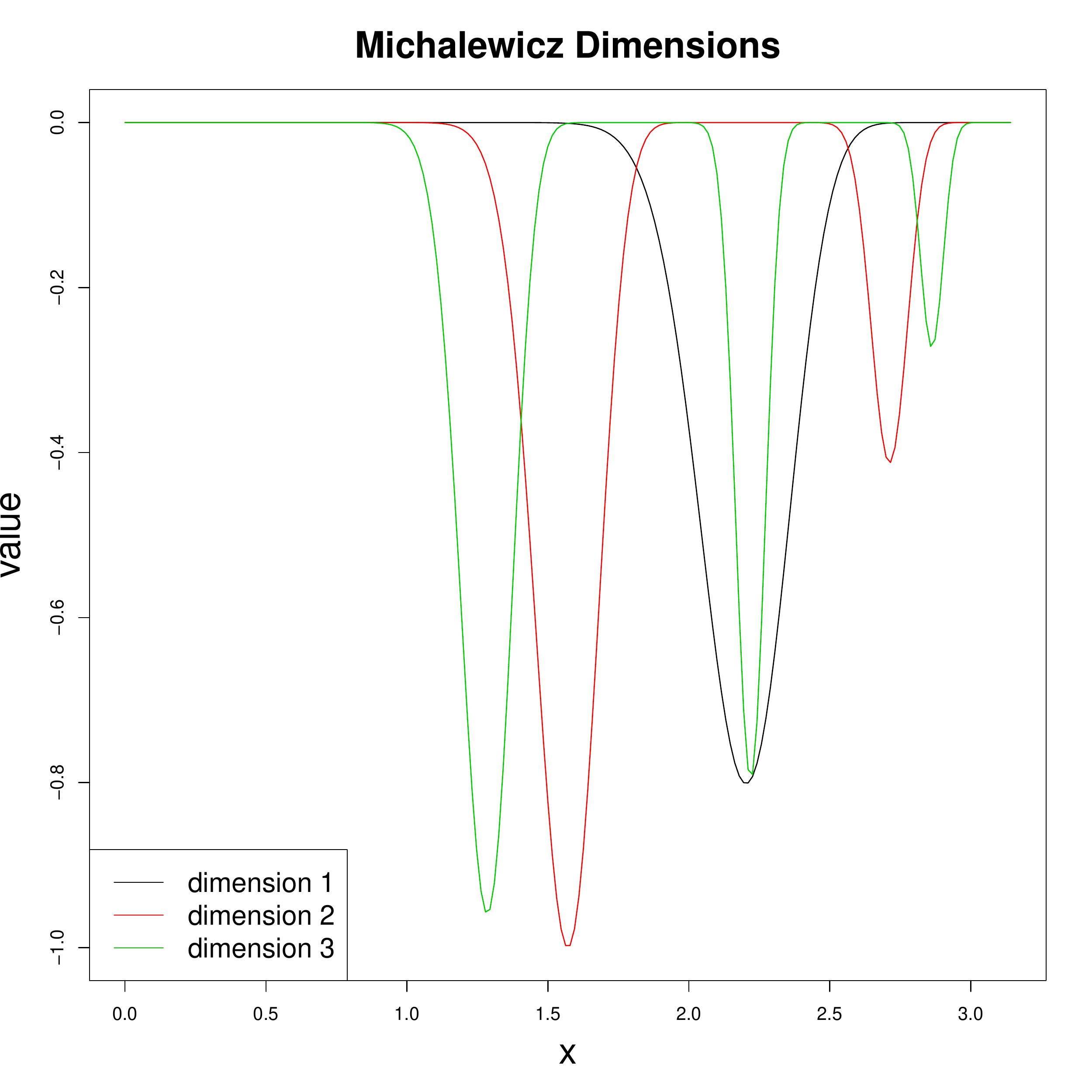}%
		\includegraphics[width=.5\textwidth]{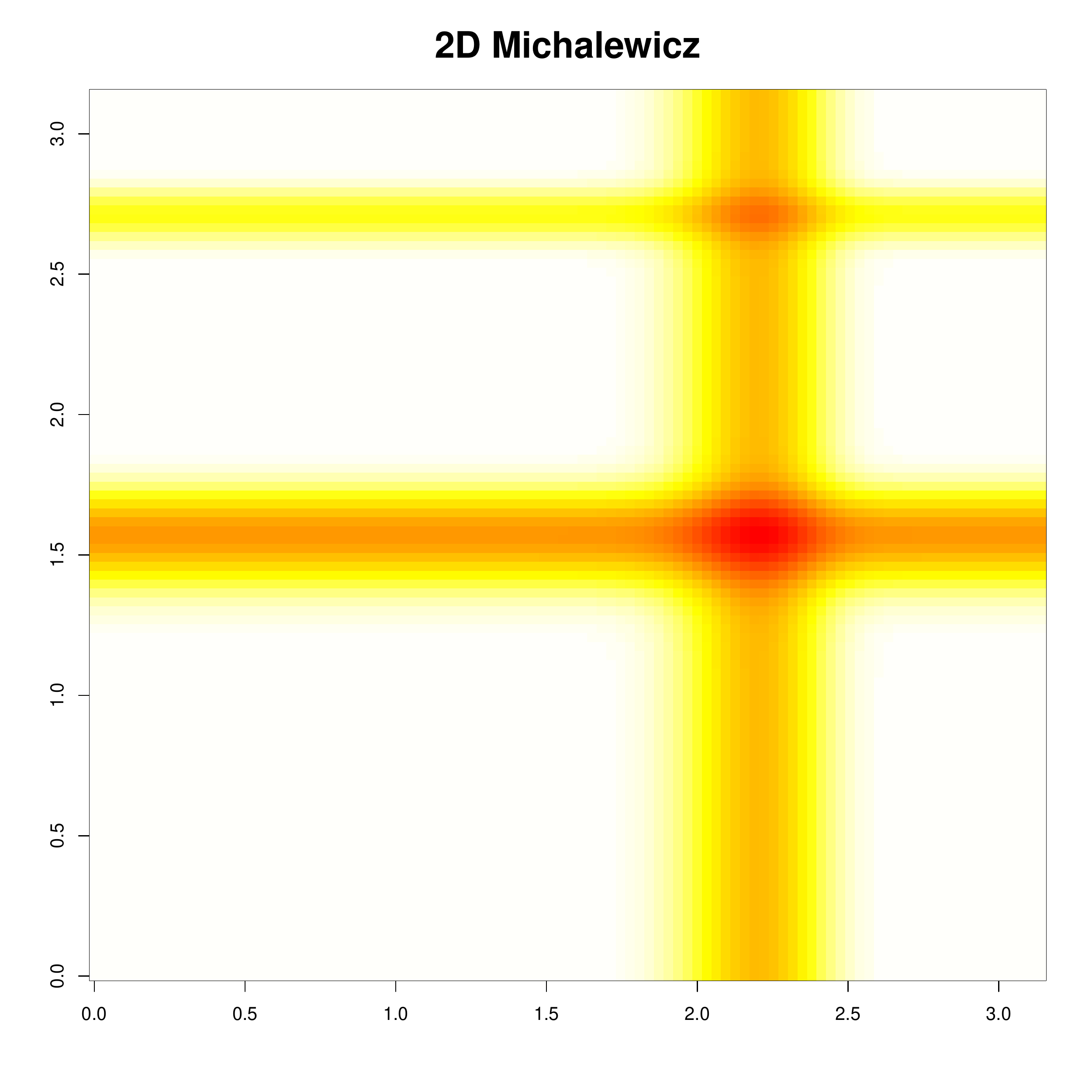}
		\vspace{-0.75cm}
		\caption{Left: The relative values of the first three dimensions of the
		Michalewicz function.  The combined 3d surface can be reconstructed by
		adding points from these lines together. Right: a visualization of the
		Michalewicz function in 2d.}
		\label{fig:michal}
	\end{figure}

	For a final illustration consider the Michalewicz function
	\citep{molga:smutnicki:2005} in 3d.  This is a good test function for PALM
	because it is highly non-stationary with large flat areas abutting drastic
	drops.  As the number of dimensions increases, the number and severity of
	the dips also increases.  For any number of dimensions, $d$, and
	steepness parameter $m$, the Michalewicz function is defined over
	$[0,\pi]^d$ as
	\[
	f(\mathbf{x}) = -\sum_{i=1}^{d} \sin(x_i) \sin^{2m}\left(\frac{ix_i^2}{\pi}\right).
	\]
	Because it is defined additively over inputs, a slice along any dimension
	will have the same shape regardless of its position in the other $d-1$
	dimensions. This allows one to intuit high dimensional dynamics visually by
	overlaying one-dimensional slices.  Figure \ref{fig:michal} shows slices
	created by the first three inputs. Any output in the 3d surface can be found
	by adding the $x$ value for the individual dimensions with the corresponding
	point along that dimension's line.  The second panel shows a heat
	plot rendering in two dimensions, combining the first two slices from the
	left panel.  The shape of the function in either direction remains the same
	regardless of the slice, although the depth of the slice changes.
	
	To test the effectiveness of space-filling versus sequential designs for
	local experts in three dimensions, we generated a grid of forty points in 3d
	from the Michalewicz function, giving a total training data size of
	$N=64,000$. We added $\mathcal{N}(0, \sigma^2 = 0.0025)$ noise to compare our
	methods in a stochastic setting.  Finally the computational resources of
	both center selection methods were restricted to budgets of $K
	\in \{100, 200\}$, separately. The testing data set, over which PALMs
	were compared on RMSE and score, was generated the same way on a grid of
	size $39^3$. The entire process, from noise generation to center location
	selection, to fitting and prediction, was repeated one hundred times to get
	an accurate estimate of the comparison through Monte Carlo averaging.

\input{resultsmichal.txt}

	Focusing first on the $K = 100$ case, Table \ref{resultsm} shows that
	despite a notable improvement on RMSE for sequential over space-filling
	center selection, score is only slightly better. This is because the
	space-filled variation yields lower average variance. Note that the
	improvement in RMSE is still enough that the model scores better despite
	having higher variance, indicating that we have spent our computational
	resources wisely.  When we increase the number of local experts to $K =
	200$, the gap in RMSE closes slightly, while the difference in score becomes
	negligible.  This confirms our instinct that sequential selection should be
	used in situations where computational resources are limited.

	\section{Discussion} \label{sec:discuss}
	
	We have introduced a framework for local model averaging of Gaussian process
	predictors as an exemplar in a potentially wide class of such methods which
	we have dubbed PALM, for precision aggregated local models.  The ingredients
	are a flexible local fit providing predictive means and variances, which we
	take from LAGP, and an aggregation scheme, which we take as inverse
	precision with adjustments for between-model covariance and overall ensemble
	size.  We get all that while remaining in the class of GP regression -- the
	resulting PALM-LAGP is a GP!  Code to reproduce all of the results and
	plots used in this paper is open source and can be found in Gramacy Lab
	LAGP git repository.
\begin{center}
\url{https://bitbucket.org/gramacylab/lagp/src/master/R/boosting}
\end{center}

	It is easy to imagine other PALM instances.  A number of other (locally
	applied) GP methods could be accommodated directly. We could swap the inverse
	squared exponential covariance function for other popular choices such as
	the Matern covariance, tapered or compactly supported covariance functions
	\citep{kaufman_covariance_2008, kaufman_efficient_2011}, or a non-stationary
	options \citep[e.g.,][]{ma_spatial_2019}.  It would not significantly change
	the method to swap out a traditional Gaussian process for inducing points
	methods \citep{snelson:ghahr:2006}. Modern sparse spatial
	processes, such as a nearest neighbor GP
	\citep[NNGP,][]{datta_hierarchical_2016, datta_non-separable_2015} can be
	swapped in.  The only significant change to the PALM framework laid out
	above would be the necessity for a new sequential selection scheme without
	the existence of a ``center'' location for the local expert.  

	Even a simple linear model can be used as the local predictor.  In the
	repository, the file {\tt lmPALM.R} contains code that runs a PALM predictor
	using a third order linear approximation locally.  The local experts are
	made by overlapping partitions of the input space for one and two
	dimensional examples: a sum of five randomly generated sine waves and
	Herbie's tooth, respectively. Covariance is determined via empirical
	predictive correlation between the points used to build two models.  These
	examples are made heteroskedastic by choosing a random point in the input
	space to be the center for radially increasing variance. Performance on both
	the mean and variance surfaces can be observed in the plots contained
	within. Additionally, the boundaries between any existing partition model,
	such as \citep{gramacy2008bayesian, park_patchwork_2017}, can be made smooth
	by applying the PALM weighting scheme.

	Other embellishments to the wider framework naturally suggest themselves.
	For example,  There is no reason why the greedy decisions of Section
	\ref{sec:seq} could not be revisited later if computational resources
	permit.  That is, we could entertain re-positioning those centers, either
	one at a time or on-mass, using the same score criterion.   One way to
	implement that idea is to select a local expert for repositioning, which
	could be implemented by removing it from the ensemble and then following the
	same greedy sequential procedure for selecting a location to add it back in.
	We have experimented extensively with this ``cycling'' procedure, with code
	provided in the git repository.  The idea was ultimately abandoned for
	presentation in the main body of this manuscript because the removal process
	necessitates forming model predictions without each of the local expert
	models.  This is an $O\left(K^3\right)$ procedure as the $O(K^2)$
	predictive process must be repeated without each of the $K$ centers.
	
	We envision a number potential opportunities for statistical and
	computational performance gains.  Exploiting independence in inferences in
	order to parallelize the computational flow is one such aspect, alluded to
	at length earlier. On statistical efficiency fonts, it may be beneficial to
	re-select the local designs for PALM's local experts after the
	hyperparameters have been learned, e.g., using maximum likelihood estimates
	from LAGP.  It may help to select a space filling design on the
	LAGP centers that is weighted in the input space based on the lengthscale
	parameters as well \citep{sun2019emulating}. Additionally, for some cases we may want to
	include variance explicitly in our sequential selection algorithm rather
	than exclusively picking center locations based on high residuals, and the
	idea that a relatively space filling point should have some marginal
	variance reducing properties.
	
	Finally, there are a number of PALM tuning parameters that we set by
	intuition, but which could potentially be optimized.  First, we selected
	$n=50$ as the number of points a local model should be trained on,
	primarily because this is the default in the {\tt laGP} package.  Within
	the provided code, this can be changed by altering the defaults to the
	{\tt aGP} function.  Of course, users should be aware of the inverse
	quadratic relationship between predictive time and accuracy based on the
	size of local GP models.  While the default we are using works well, it
	may be that other local model sizes may result in a better global model.
	One could imagine choosing such a tuning parameter via cross validation
	(CV).  It may also possible to incorporate varying local model sizes in
	different areas of the input space that are easier or more difficult to
	capture.  Updating local model size may be an alternative to sequential
	selection that allows the model to learn the relative complexity of the
	input space.  Selection of the power to which weights are raised is
	another such tuning parameter.  We chose to use $\log_d K$ as a slowly
	increasing function of the number of centers modulated for dimension of
	the input space.  CV may well be a more data-centric option for that
	parameter as well.  The LAGP repository provides code that that optimizes
	the predictive power for a specific model, but it runs slowly and offers
	little benefit to statistical efficiency.

	\bibliography{Biblio} \bibliographystyle{jasa}

\end{document}

%% file: results.txt
\begin{table}[ht]
\centering
\begin{tabular}{rrrr}
  \hline
 & RMSE & Score & Time \\ 
  \hline
LAGP & 0.0515 & 4.9062 & 0.0628 \\ 
  PALM & 0.0525 & 4.8887 & 0.0021 \\ 
   \hline
\end{tabular}
\caption{Mean RMSE, score, and predictive time per point on Herbie's
tooth.  PALM had an average model build time of 93 seconds, for a 114 seconds to predict 9801 points.}
\label{results}
\end{table}

%% file: temps.txt

\begin{table}[ht]
\centering
\begin{tabular}{rrrrrrrrr}
  \hline
 & MAE & RMSE & CRPS & INT & CVG & Time & Cores & Score \\ 
  \hline
FRK & 1.96 & 2.44 & 1.44 & 14.08 & 0.79 & 2.32 & 1 &  \\ 
  Gapfill & 1.33 & 1.86 & 1.17 & 34.78 & 0.36 & 1.39 & 40 &  \\ 
  Lattice Krig & 1.22 & 1.68 & 0.87 & 7.55 & 0.96 & 27.92 & 1 &  \\ 
  Metakriging & 2.08 & 2.50 & 1.44 & 10.77 & 0.89 & 2888.52 & 30 &  \\ 
  MRA & 1.33 & 1.85 & 0.94 & 8.00 & 0.92 & 15.61 & 1 &  \\ 
  NNGP Conjugate & 1.21 & 1.64 & 0.85 & 7.57 & 0.95 & 2.06 & 10 &  \\ 
  NNGP Response & 1.24 & 1.68 & 0.87 & 7.50 & 0.94 & 42.85 & 10 &  \\ 
  Partition & 1.41 & 1.80 & 1.02 & 10.49 & 0.86 & 79.98 & 55 &  \\ 
  Pred. Proc. & 2.05 & 2.52 & 1.85 & 26.24 & 0.75 & 640.48 & 1 &  \\ 
  SPDE & 1.10 & 1.53 & 0.83 & 8.85 & 0.97 & 120.33 & 2 &  \\ 
  Tapering & 1.87 & 2.45 & 1.32 & 10.31 & 0.93 & 133.26 & 1 &  \\ 
  Periodic Embedding & 1.29 & 1.79 & 0.91 & 7.44 & 0.93 & 9.81 & 1 &  \\ 
  LAGP & 1.65 & 2.08 & 1.17 & 10.81 & 0.83 & 2.27 & 40 & -2.55 \\ 
  PALM & 1.59 & 1.93 & 1.15 & 11.78 & 0.78 & 4.64 & 1 & -2.85 \\ 
  Global + PALM & 1.44 & 1.76 & 1.03 & 9.28 & 0.84 & 4.64 & 1 & -2.39 \\ 
   \hline
\end{tabular}
\caption{The original table from Heaton's bake-off paper with added rows for PALM comparison, and an added column for the proper scoring rule} \label{t:sat}
\end{table}

%% file: resultsmichal.txt
\begin{table}[ht]
\centering
\begin{tabular}{|r|rr|rr|}
  \hline
  & \multicolumn{2}{|c|}{100 Centers} & \multicolumn{2}{|c|}{200 Centers}\\
  \hline
  \hline
 & RMSE & Score & RMSE & Score\\ 
  \hline
Spacefill & 0.1237 & 3.5623 & 0.0878 & 4.1012 \\ 
  Sequential & 0.1073 & 3.5704 & 0.0776 & 4.1212 \\ 
   \hline
\end{tabular}
\caption{Average RMSE and Score for spacefilling models on 3d Michalewicz.} 
\label{resultsm}
\end{table}

%% file: PALM_ArXiv.bbl
\begin{thebibliography}{47}
\newcommand{\enquote}[1]{``#1''}
\expandafter\ifx\csname natexlab\endcsname\relax\def\natexlab#1{#1}\fi

\bibitem[\protect\citename{Binois et~al., }2018]{binois2018practical}
Binois, M., Gramacy, R.~B., and Ludkovski, M. (2018).
\newblock \enquote{Practical heteroscedastic gaussian process modeling for
  large simulation experiments.}
\newblock {\em Journal of Computational and Graphical Statistics\/}, 27, 4,
  808--821.

\bibitem[\protect\citename{Chen and Ren, }2009]{chen_bagging_2009}
Chen, T. and Ren, J. (2009).
\newblock \enquote{Bagging for {Gaussian} process regression.}
\newblock {\em Neurocomputing\/}, 72, 7, 1605--1610.

\bibitem[\protect\citename{Chipman et~al., }2010]{chipman2010bart}
Chipman, H., George, E., and McCulloch, R. (2010).
\newblock \enquote{{BART}: {B}ayesian additive regression trees.}
\newblock {\em The Annals of Applied Statistics\/}, 4, 1, 266--298.

\bibitem[\protect\citename{Chipman et~al., }2012]{chipman2012sequential}
Chipman, H., Ranjan, P., and Wang, W. (2012).
\newblock \enquote{Sequential design for computer experiments with a flexible
  {B}ayesian additive model.}
\newblock {\em Canadian Journal of Statistics\/}, 40, 4, 663--678.

\bibitem[\protect\citename{Cochran, }1954]{cochran_combination_1954}
Cochran, W.~G. (1954).
\newblock \enquote{The {Combination} of {Estimates} from {Different}
  {Experiments}.}
\newblock {\em Biometrics\/}, 10, 1, 101--129.

\bibitem[\protect\citename{Cohn, }1994]{cohn1994neural}
Cohn, D. (1994).
\newblock \enquote{Neural network exploration using optimal experiment design.}
\newblock In {\em Advances in neural information processing systems\/},
  679--686.

\bibitem[\protect\citename{Cressie, }1991]{cressie:1993}
Cressie, N. (1991).
\newblock {\em Statistics for Spatial Data, {\em revised edition}\/}.
\newblock John Wiley and Sons, Inc.

\bibitem[\protect\citename{Cressie and Johannesson, }2008]{cressie:joh:2008}
Cressie, N. and Johannesson, G. (2008).
\newblock \enquote{Fixed Rank Kriging for Very Large Data Sets.}
\newblock {\em Journal of the Royal Statistical Soceity, Series B\/}, 70, 1,
  209--226.

\bibitem[\protect\citename{Datta et~al., }2016]{datta_hierarchical_2016}
Datta, A., Banerjee, S., Finley, A.~O., and Gelfand, A.~E. (2016).
\newblock \enquote{Hierarchical {Nearest}-{Neighbor} {Gaussian} {Process}
  {Models} for {Large} {Geostatistical} {Datasets}.}
\newblock {\em Journal of the American Statistical Association\/}, 111, 514,
  800--812.

\bibitem[\protect\citename{Datta et~al., }2015]{datta_non-separable_2015}
Datta, A., Banerjee, S., Finley, A.~O., Hamm, N. A.~S., and Schaap, M. (2015).
\newblock \enquote{Non-separable {Dynamic} {Nearest}-{Neighbor} {Gaussian}
  {Process} {Models} for {Large} spatio-temporal {Data} {With} an {Application}
  to {Particulate} {Matter} {Analysis}.}
\newblock {\em arXiv:1510.07130 [stat]\/}.
\newblock ArXiv: 1510.07130.

\bibitem[\protect\citename{Furrer et~al., }2006]{furrer:genton:nychka:2006}
Furrer, R., Genton, M., and Nychka, D. (2006).
\newblock \enquote{Covariance Tapering for Interpolation of Large Spatial
  Datasets.}
\newblock {\em Journal of Computational and Graphical Statistics\/}, 15,
  502--523.

\bibitem[\protect\citename{Gneiting and Raftery, }2007]{gneiting2007scoring}
Gneiting, T. and Raftery, A. (2007).
\newblock \enquote{Strictly Proper Scoring Rules, Prediction, and Estimation.}
\newblock {\em Journal of the American Statistical Association\/}, 102, 477,
  359--378.

\bibitem[\protect\citename{Goldberg et~al.,
  }1998]{goldberg:williams:bishop:1998}
Goldberg, P.~W., Williams, C.~K., and Bishop, C.~M. (1998).
\newblock \enquote{Regression with input-dependent noise: A {G}aussian process
  treatment.}
\newblock In {\em Advances in Neural Information Processing Systems\/},
  vol.~10,  493--499. Cambridge, MA: MIT press.

\bibitem[\protect\citename{Gramacy and Apley, }2015]{gramacy2015local}
Gramacy, R. and Apley, D. (2015).
\newblock \enquote{Local Gaussian process approximation for large computer
  experiments.}
\newblock {\em Journal of Computational and Graphical Statistics\/}, 24, 2,
  561--578.

\bibitem[\protect\citename{Gramacy and Lee, }2008]{gramacy2008bayesian}
Gramacy, R. and Lee, H. (2008).
\newblock \enquote{Bayesian treed Gaussian process models with an application
  to computer modeling.}
\newblock {\em Journal of the American Statistical Association\/}, 103, 483,
  1119--1130.

\bibitem[\protect\citename{Gramacy et~al., }2014]{gramacy2014massively}
Gramacy, R., Niemi, J., and Weiss, R. (2014).
\newblock \enquote{Massively parallel approximate {G}aussian process
  regression.}
\newblock {\em SIAM/ASA Journal on Uncertainty Quantification\/}, 2, 1,
  564--584.

\bibitem[\protect\citename{Gramacy and Polson, }2011]{gramacy:polson:2011}
Gramacy, R. and Polson, N. (2011).
\newblock \enquote{Particle Learning of Gaussian Process Models for Sequential
  Design and Optimization.}
\newblock {\em Journal of Computational and Graphical Statistics\/}, 20, 1,
  102--118.

\bibitem[\protect\citename{Gramacy, }2016]{gramacy_lagp:_2016}
Gramacy, R.~B. (2016).
\newblock \enquote{{laGP}: {Large}-{Scale} {Spatial} {Modeling} via {Local}
  {Approximate} {Gaussian} {Processes} in {R}.}
\newblock {\em Journal of Statistical Software\/}, 72, 1.

\bibitem[\protect\citename{Gramacy, }2020]{gramacy2020surrogates}
--- (2020).
\newblock {\em Surrogates: {G}aussian Process Modeling, Design and Optimization
  for the Applied Sciences\/}.
\newblock Boca Raton, Florida: Chapman Hall/CRC.
\newblock \url{http://bobby.gramacy.com/surrogates/}.

\bibitem[\protect\citename{Haaland and Qian, }2011]{haaland:qian:2012}
Haaland, B. and Qian, P. (2011).
\newblock \enquote{Accurate Emulators for Large-Scale Computer Experiments.}
\newblock {\em Annals of Statistics\/}, 39, 6, 2974--3002.

\bibitem[\protect\citename{Heaton et~al., }2019]{heaton2019case}
Heaton, M.~J., Datta, A., Finley, A.~O., Furrer, R., Guinness, J., Guhaniyogi,
  R., Gerber, F., Gramacy, R.~B., Hammerling, D., Katzfuss, M., et~al. (2019).
\newblock \enquote{A case study competition among methods for analyzing large
  spatial data.}
\newblock {\em Journal of Agricultural, Biological and Environmental
  Statistics\/}, 24, 3, 398--425.

\bibitem[\protect\citename{Johnson et~al., }1990]{johnson1990minimax}
Johnson, M., Moore, L., and Ylvisaker, D. (1990).
\newblock \enquote{Minimax and maximin distance designs.}
\newblock {\em Journal of statistical planning and inference\/}, 26, 2,
  131--148.

\bibitem[\protect\citename{Katzfuss, }2017]{katzfuss2017multi}
Katzfuss, M. (2017).
\newblock \enquote{A multi-resolution approximation for massive spatial
  datasets.}
\newblock {\em Journal of the American Statistical Association\/}, 112, 517,
  201--214.

\bibitem[\protect\citename{Katzfuss and Guinness, }2018]{katzfuss2018general}
Katzfuss, M. and Guinness, J. (2018).
\newblock \enquote{A general framework for Vecchia approximations of {G}aussian
  processes.}
\newblock {\em arXiv preprint arXiv:1708.06302\/}.

\bibitem[\protect\citename{Kaufman et~al., }2012]{kaufman:etal:2012}
Kaufman, C., Bingham, D., Habib, S., Heitmann, K., and Frieman, J. (2012).
\newblock \enquote{Efficient Emulators of Computer Experiments using Compactly
  Supported Correlation Functions, with An Application to Cosmology.}
\newblock {\em The Annals of Applied Statistics\/}, 5, 4, 2470--2492.

\bibitem[\protect\citename{Kaufman et~al., }2011]{kaufman_efficient_2011}
Kaufman, C.~G., Bingham, D., Habib, S., Heitmann, K., and Frieman, J.~A.
  (2011).
\newblock \enquote{Efficient emulators of computer experiments using compactly
  supported correlation functions, with an application to cosmology.}
\newblock {\em The Annals of Applied Statistics\/}, 5, 4, 2470--2492.
\newblock ArXiv: 1107.0749.

\bibitem[\protect\citename{Kaufman et~al., }2008]{kaufman_covariance_2008}
Kaufman, C.~G., Schervish, M.~J., and Nychka, D.~W. (2008).
\newblock \enquote{Covariance {Tapering} for {Likelihood}-{Based} {Estimation}
  in {Large} {Spatial} {Data} {Sets}.}
\newblock {\em Journal of the American Statistical Association\/}, 103, 484,
  1545--1555.

\bibitem[\protect\citename{Kim et~al., }2005]{kim2005analyzing}
Kim, H., Mallick, B., and Holmes, C. (2005).
\newblock \enquote{Analyzing nonstationary spatial data using piecewise
  {G}aussian processes.}
\newblock {\em Journal of the American Statistical Association\/}, 100, 470,
  653--668.

\bibitem[\protect\citename{Lee et~al., }2011]{lee2011optimization}
Lee, H., Gramacy, R., Linkletter, C., and Gray, G. (2011).
\newblock \enquote{Optimization subject to hidden constraints via statistical
  emulation.}
\newblock {\em Pacific Journal of Optimization\/}, 7, 3, 467--478.

\bibitem[\protect\citename{Lloyd, }1982]{lloyd1982kmeans}
Lloyd, S.~P. (1982).
\newblock \enquote{Least squares quantization in PCM.}
\newblock {\em IEEE Transactions on Information Theory\/}, 28, 2, 129--137.

\bibitem[\protect\citename{Ma et~al., }2019]{ma_spatial_2019}
Ma, P., Kang, E.~L., Braverman, A., and Nguyan, H. (2019).
\newblock \enquote{Spatial {Statistical} {Downscaling} for {Constructing}
  {High}-{Resolution} {Nature} {Runs} in {Global} {Observing} {System}
  {Simulation} {Experiments}.}
\newblock {\em Technometrics\/}, 61, 3, 322--340.

\bibitem[\protect\citename{Molga and Smutnicki, }2005]{molga:smutnicki:2005}
Molga, M. and Smutnicki, C. (2005).
\newblock \enquote{Test Functions for Optimization Needs.}

\bibitem[\protect\citename{Neal, }1998]{neal1998regression}
Neal, R. (1998).
\newblock \enquote{Regression and classification using {G}aussian process
  priors.}
\newblock {\em Bayesian Statistics\/}, 6, 475.

\bibitem[\protect\citename{Nychka et~al., }2002]{nychka:wykle:royle:2002}
Nychka, D., Wikle, C., and Royle, J. (2002).
\newblock \enquote{Multiresolution Models for Nonstationary Spatial Covariance
  Functions.}
\newblock {\em Statistical Modelling\/}, 2, 315--331.

\bibitem[\protect\citename{Park and Apley, }2017]{park_patchwork_2017}
Park, C. and Apley, D. (2017).
\newblock \enquote{Patchwork {Kriging} for {Large}-scale {Gaussian} {Process}
  {Regression}.}
\newblock {\em arXiv:1701.06655 [cs, stat]\/}.
\newblock ArXiv: 1701.06655.

\bibitem[\protect\citename{Park and Huang, }2016]{park_efficient_2016}
Park, C. and Huang, J.~Z. (2016).
\newblock \enquote{Efficient {Computation} of {Gaussian} {Process} {Regression}
  for {Large} {Spatial} {Data} {Sets} by {Patching} {Local} {Gaussian}
  {Processes}.}
\newblock {\em Journal of Machine Learning Research\/}, 17, 174, 1--29.

\bibitem[\protect\citename{Park et~al., }2011]{park_domain_2011}
Park, C., Huang, J.~Z., and Ding, Y. (2011).
\newblock \enquote{Domain {Decomposition} {Approach} for {Fast} {Gaussian}
  {Process} {Regression} of {Large} {Spatial} {Data} {Sets}.}
\newblock {\em Journal of Machine Learning Research\/}, 12, May, 1697--1728.

\bibitem[\protect\citename{{Qui\~nonero--Candela} and Rasmussen,
  }2005]{qc:rasmu:2005}
{Qui\~nonero--Candela}, J. and Rasmussen, C. (2005).
\newblock \enquote{A Unifying View of Sparse Approximate Gaussian Process
  Regression.}
\newblock {\em Journal of Machine Learning Research\/}, 6, 1939--1959.

\bibitem[\protect\citename{Rasmussen and Williams, }2006]{rasmu:will:2006}
Rasmussen, C.~E. and Williams, C. K.~I. (2006).
\newblock {\em Gaussian Processes for Machine Learning\/}.
\newblock The MIT Press.

\bibitem[\protect\citename{Rushdi et~al., }2016]{rushdi_vps:_2016}
Rushdi, A., P.~Swiler, L., T.~Phipps, E., D'Elia, M., and Ebeida, M. (2016).
\newblock \enquote{{VPS}: {Voronoi} {Piecewise} {Surrogate} {Models} for
  {High}-{Dimensional} {Data} {Fitting}.}
\newblock {\em International Journal for Uncertainty Quantification\/}, 7.

\bibitem[\protect\citename{Sacks et~al., }1989]{sack:welc:mitc:wynn:1989}
Sacks, J., Welch, W.~J., Mitchell, T.~J., and Wynn, H.~P. (1989).
\newblock \enquote{Design and Analysis of Computer Experiments.}
\newblock {\em Statistical Science\/}, 4, 409--435.

\bibitem[\protect\citename{Sang and Huang, }2012]{sang:huang:2012}
Sang, H. and Huang, J.~Z. (2012).
\newblock \enquote{A Full Scale Approximation of Covariance Functions for Large
  Spatial Data Sets.}
\newblock {\em Journal of the Royal Statistical Society: Series B\/}, 74, 1,
  111--132.

\bibitem[\protect\citename{Santner et~al., }2018]{santner2018design}
Santner, T., Williams, B., and Notz, W. (2018).
\newblock {\em The Design and Analysis of Computer Experiments, Second
  Edition\/}.
\newblock New York, NY: Springer--Verlag.

\bibitem[\protect\citename{Snelson and Ghahramani, }2006]{snelson:ghahr:2006}
Snelson, E. and Ghahramani, Z. (2006).
\newblock \enquote{Sparse Gaussian Processes using Pseudo-inputs.}
\newblock In {\em Advances in Neural Information Processing Systems\/},
  1257--1264. MIT press.

\bibitem[\protect\citename{Sun et~al., }2019]{sun2019emulating}
Sun, F., Gramacy, R.~B., Haaland, B., Lawrence, E., and Walker, A. (2019).
\newblock \enquote{Emulating satellite drag from large simulation experiments.}
\newblock {\em SIAM/ASA Journal on Uncertainty Quantification\/}, 7, 2,
  720--759.

\bibitem[\protect\citename{Tresp, }2000]{tresp_bayesian_2000}
Tresp, V. (2000).
\newblock \enquote{A {Bayesian} {Committee} {Machine}.}
\newblock {\em Neural Computation\/}, 12, 11, 2719--2741.

\bibitem[\protect\citename{Vapnik, }2013]{vapnik2013nature}
Vapnik, V. (2013).
\newblock {\em The nature of statistical learning theory\/}.
\newblock Springer science \& business media.

\end{thebibliography}
